\documentclass[10pt,a4paper]{article}

\usepackage[utf8]{inputenc}
\usepackage[T1]{fontenc}
\usepackage{authblk}
\usepackage{lmodern}
\usepackage{microtype}
\usepackage[a4paper,top=20mm,bottom=23mm,left=22mm,right=22mm,headheight=15pt,footskip=11mm]{geometry}
\usepackage{graphicx}
\usepackage{tikz}
\usetikzlibrary{arrows.meta}
\usepackage[dvipsnames]{xcolor}
\usepackage{fancyhdr}
\usepackage{titlesec}
\usepackage{needspace}
\usepackage{amsmath,amssymb,amsthm}
\usepackage[inline,shortlabels]{enumitem}
\usepackage{booktabs}
\usepackage{array}
\usepackage{multirow}
\usepackage{tabularx}
\usepackage{float}
\usepackage{pifont}
\usepackage{listings}
\usepackage[skins,breakable,listings]{tcolorbox}
\usepackage{varwidth}
\usepackage{hyperref}
\hypersetup{colorlinks=true,citecolor=blue,linkcolor=blue,urlcolor=blue}
\usepackage[numbers]{natbib}
\usepackage{soul}

\usepackage{bm}
\usepackage{relsize}

\definecolor{QSTInk}{RGB}{31,40,54}
\definecolor{QSTNavy}{RGB}{24,55,91}
\definecolor{QSTMuted}{RGB}{102,116,135}
\definecolor{QSTPanelBlueDark}{RGB}{21,64,107}
\definecolor{QSTPanelBlueFrame}{RGB}{92,139,181}
\definecolor{QudeBlue}{RGB}{34,55,199}
\definecolor{QudeViolet}{RGB}{108,49,225}

\definecolor{QITLeanDecl}{RGB}{15,77,151}
\definecolor{QITLeanNamespace}{RGB}{16,108,137}
\definecolor{QITLeanProof}{RGB}{49,118,89}
\definecolor{QITLeanType}{RGB}{36,97,153}
\definecolor{QITLeanDomain}{RGB}{38,83,128}
\definecolor{QITLeanComment}{RGB}{94,111,124}
\definecolor{QITLeanString}{RGB}{26,118,116}
\definecolor{QITLeanWarn}{RGB}{145,69,60}

\DeclareRobustCommand{\PaperShortTitle}{QECert}
\newcommand{\QudeLeapLogo}[1]{\includegraphics[#1]{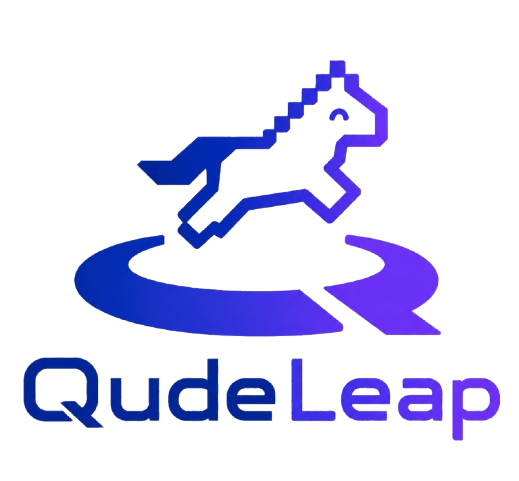}}

\titleformat{\section}
  {\Large\sffamily\bfseries\color{QSTPanelBlueDark}}
  {\thesection}{0.65em}{}
  [\vspace{0.15em}\color{QSTPanelBlueFrame!55}\titlerule]
\titleformat{\subsection}
  {\large\sffamily\bfseries\color{QSTNavy}}
  {\thesubsection}{0.65em}{}
\titleformat{\subsubsection}
  {\normalsize\sffamily\bfseries\color{QSTInk}}
  {\thesubsubsection}{0.65em}{}

\makeatletter
\renewcommand{\maketitle}{\thispagestyle{empty}\begingroup
  \setlength{\parindent}{0pt}\noindent
  \begin{minipage}[t]{0.8\textwidth}
    \vspace{0pt}{\raggedright\hyphenpenalty=10000\exhyphenpenalty=10000
      \sffamily\bfseries\color{QSTInk}\fontsize{23.5}{27.5}\selectfont
      \@title\par}\end{minipage}\hfill
  \begin{minipage}[t]{0.18\textwidth}
    \vspace{0pt}\raggedleft
    \QudeLeapLogo{width=2.3cm}\end{minipage}
  \vspace{4mm}

  \noindent
  {\color{QudeBlue}\rule{0.22\textwidth}{1.8pt}}{\color{QudeViolet}\rule{0.78\textwidth}{1.8pt}}\par
  \vspace{5mm}

  \begin{center}
    {\sffamily\normalsize\color{QSTInk}\@author\par}\end{center}
  \vspace{3mm}
  \@thanks
  \endgroup
  \setcounter{footnote}{0}}
\makeatother

\lstdefinelanguage{Lean4}{
  sensitive=true,
  alsoletter={_},
  morekeywords=[1]{
    abbrev,class,def,deriving,example,inductive,instance,lemma,
    noncomputable,opaque,structure,theorem
  },
  morekeywords=[2]{
    end,export,import,module,namespace,open,private,protected,public,section,
    variable,variables,where
  },
  morekeywords=[3]{
    by,calc,change,do,else,exact,exists,for,forall,from,fun,have,if,
    inferInstance,infix,infixl,infixr,intro,intros,let,letI,macro,match,
    obtain,rcases,refine,rfl,rw,show,syntax,termination_by,then,with
  },
  morekeywords=[4]{
    AddCommGroup,Basis,Bool,Complex,DecidableEq,ENNReal,Equiv,Fact,False,
    Field,Fin,Finset,Fintype,Group,InnerProductSpace,Int,List,Matrix,
    Module,MonoidAlgebra,Nat,NeZero,None,Nonempty,Nontrivial,
    NormedAddCommGroup,Option,Pi,PMF,Prop,Quotient,Real,Semiring,Set,
    Some,Sort,String,Submodule,Subtype,Sum,True,Type,Unit,ZMod
  },
  morekeywords=[5]{axiom,sorry,admit},
  morekeywords=[6]{
    BalancedProduct,BalancedQubitIndex,BH12PathScheduleIndex,Binary,
    BPTCommutingProjectorModel,CayleySquare,ChainComplexData,
    CheckOrbit401,Coding,CompatibleDegreePrime,Cyclic,CyclicGroup,
    Data,DensityOperator,DependentBatch,EdgeOrbit401,EncodedQubitParameter,
    ErrorParameter,Family,FamilyQubitIndex,FiniteKrausFamily,
    FiniteQuantumSystem,FiniteQuditCheckPresentation,
    HorizontalSubsystem,InputErrorModel,LeftRightCayleySquareComplex,
    LiftedProduct,LinearCode,LPSVertexGroup,MembraneSector,OddRow,
    OneQubit,Operator,PauliErrorPath,PauliLikePair,PK22Check,PK22Direction,
    Plaquette,QEC,QLDPC,QuantumChannel,QuantumCode,QuantumCodeFamily,
    Qubit,QubitIndex,QubitSpace,Qudit,QuditRegister,QuditSubsystem,
    RawData,Semantics,SmallSetExpansion,SourceValidBinary,ThermalEncoding,
    Why24LocalStochasticPreparation,Word
  },
  morecomment=[l]{--},
  morecomment=[s]{/-}{-/},
  morestring=[b]",
}

\lstdefinestyle{QudeLeanListing}{
  language=Lean4,
  basicstyle=\fontsize{9.8}{11.5}\selectfont\ttfamily\color{QSTInk},
  keywordstyle=[1]\color{QITLeanDecl}\bfseries,
  keywordstyle=[2]\color{QITLeanNamespace}\bfseries,
  keywordstyle=[3]\color{QITLeanProof},
  keywordstyle=[4]\color{QITLeanType}\bfseries,
  keywordstyle=[5]\color{QITLeanWarn}\bfseries,
  keywordstyle=[6]\color{QITLeanDomain}\bfseries,
  commentstyle=\color{QITLeanComment},
  stringstyle=\color{QITLeanString},
  numbers=none,
  xleftmargin=0pt,
  framexleftmargin=0pt,
  columns=fullflexible,
  keepspaces=true,
  showstringspaces=false,
  tabsize=2,
  breaklines=true,
  breakatwhitespace=false,
  breakindent=1.15em,
  breakautoindent=true,
  aboveskip=0pt,
  belowskip=0pt,
  literate=
    {_}{{\textunderscore\allowbreak}}1
    {∀}{{\ensuremath{\forall}}}1
    {∃}{{\ensuremath{\exists}}}1
    {→}{{\ensuremath{\to}}}1
    {←}{{\ensuremath{\leftarrow}}}1
    {↔}{{\ensuremath{\leftrightarrow}}}1
    {∧}{{\ensuremath{\wedge}}}1
    {∨}{{\ensuremath{\vee}}}1
    {¬}{{\ensuremath{\neg}}}1
    {≤}{{\ensuremath{\leq}}}1
    {≥}{{\ensuremath{\geq}}}1
    {≠}{{\ensuremath{\neq}}}1
    {∈}{{\ensuremath{\in}}}1
    {⟨}{{\ensuremath{\langle}}}1
    {⟩}{{\ensuremath{\rangle}}}1
    {λ}{{\ensuremath{\lambda}}}1
    {α}{{\ensuremath{\alpha}}}1
    {β}{{\ensuremath{\beta}}}1
    {θ}{{\ensuremath{\theta}}}1
    {η}{{\ensuremath{\eta}}}1
    {ψ}{{\ensuremath{\psi}}}1
    {ρ}{{\ensuremath{\rho}}}1
    {ℕ}{{\ensuremath{\mathbb{N}}}}1
    {ℤ}{{\ensuremath{\mathbb{Z}}}}1
    {ℚ}{{\ensuremath{\mathbb{Q}}}}1
    {ℝ}{{\ensuremath{\mathbb{R}}}}1
    {ℂ}{{\ensuremath{\mathbb{C}}}}1
    {γ}{{\ensuremath{\gamma}}}1
    {ι}{{\ensuremath{\iota}}}1
    {κ}{{\ensuremath{\kappa}}}1
    {₀}{{\ensuremath{_0}}}1
    {₁}{{\ensuremath{_1}}}1
    {₂}{{\ensuremath{_2}}}1
    {ₗ}{{\ensuremath{_{\!l}}}}1
    {↦}{{\ensuremath{\mapsto}}}1
    {≃}{{\ensuremath{\simeq}}}1
    {×}{{\ensuremath{\times}}}1
    {ˣ}{{\ensuremath{^{\times}}}}1
    {⁻¹}{{\ensuremath{^{-1}}}}2
    {‖}{{\ensuremath{\|}}}1
    {∑}{{\ensuremath{\sum}}}1
    {·}{{\ensuremath{\cdot}}}1
    {⦃}{{\ensuremath{\{\!\!\{}}}1
    {⦄}{{\ensuremath{\}\!\!\}}}}1
    {∘}{{\ensuremath{\circ}}}1
    {⊗}{{\ensuremath{\otimes}}}1
    {ε}{{\ensuremath{\varepsilon}}}1
    {π}{{\ensuremath{\pi}}}1
}

\newtcblisting[auto counter]{leanlisting}[3][]{enhanced,
  breakable,
  lines before break=6,
  listing only,
  listing engine=listings,
  listing options={style=QudeLeanListing},
  colback=gray!5,
  colframe=gray!55,
  boxrule=0.4pt,
  arc=0.6mm,
  left=4pt,
  right=4pt,
  top=4pt,
  bottom=4pt,
  title={Listing~\thetcbcounter: #2},
  fonttitle=\sffamily\small,
  coltitle=black,
  attach boxed title to top left={yshift=-2mm,xshift=4mm},
  varwidth boxed title=\dimexpr\linewidth-8mm\relax,
  boxed title style={colback=gray!18,sharp corners,boxrule=0pt,boxsep=2pt},
  label={#3},
  before skip=0.6\baselineskip,
  after skip=0.65\baselineskip,
  #1
}

\newtcbinputlisting[use counter from=leanlisting]{\leaninputlisting}[3]{enhanced,
  breakable,
  lines before break=6,
  listing only,
  listing engine=listings,
  listing file={#3},
  listing options={style=QudeLeanListing},
  colback=gray!5,
  colframe=gray!55,
  boxrule=0.4pt,
  arc=0.6mm,
  left=4pt,
  right=4pt,
  top=4pt,
  bottom=4pt,
  title={Listing~\thetcbcounter: #1},
  title after break={Listing~\thetcbcounter\ (continued)},
  fonttitle=\sffamily\small,
  coltitle=black,
  attach boxed title to top left={yshift=-2mm,xshift=4mm},
  varwidth boxed title=\dimexpr\linewidth-8mm\relax,
  boxed title style={colback=gray!18,sharp corners,boxrule=0pt,boxsep=2pt},
  label={#2},
  before skip=0.6\baselineskip,
  after skip=0.65\baselineskip
}

\newcommand{\lean}[1]{\texttt{\detokenize{#1}}}

\theoremstyle{plain}
\newtheorem{theorem}{Theorem}[section]
\theoremstyle{definition}
\newtheorem{definition}[theorem]{Definition}

\title{QECert: Formal Foundations for Quantum Error-Correction Theory in Lean}
\author{Keming He\textsuperscript{1,2,*},
  Jiahan Chen\textsuperscript{1,2,\ensuremath{\dagger}},
  Enji Xiong\textsuperscript{2},
  Zhongyi Ni\textsuperscript{1,2},
  Lei Zhang\textsuperscript{1,2},
  Xin Wang\textsuperscript{2,\ensuremath{\ddagger}}\\
  {\small\textsuperscript{1}QudeLeap Research, Shanghai 200030, China}\\
  {\small\textsuperscript{2}The Hong Kong University of Science and Technology (Guangzhou), Guangdong 511453, China}}
\date{}

\ifdefined\hypersetup
\hypersetup{pdfcreator={},pdfproducer={}}
\else
\ifdefined{/Creator () /Producer ()}\fi
\fi
\ifdefined\pdfinfoomitdate\pdfinfoomitdate=1\fi
\ifdefined\pdfsuppressptexinfo\pdfsuppressptexinfo=-1\fi
\ifdefined\pdftrailerid\pdftrailerid{}\fi
\begin{document}
\maketitle
\begingroup
\renewcommand{\thefootnote}{\fnsymbol{footnote}}
\footnotetext[1]{keming.he2020@gmail.com}
\footnotetext[2]{jiahanchen527@gmail.com}
\footnotetext[3]{felixxinwang@hkust-gz.edu.cn}
\endgroup

\vspace{-5mm}
\begin{center}
  \small\sffamily\color{QSTMuted}\raisebox{-0.5ex}{\QudeLeapLogo{height=3.2mm}}\hspace{2pt}\href{https://github.com/QudeLeap/Lean-QECert}{github.com/QudeLeap/Lean-QECert}\end{center}

\begin{abstract}
Quantum error correction provides the mathematical basis for protecting quantum information against noise.
Existing formalizations have established foundations for stabilizer codes and code verification.
However, a systematic formal treatment of the asymptotic constructions and fundamental limitations that underpin scalable quantum error correction is still needed.
We present QECert, a Lean library providing reusable formal foundations for these theoretical questions.
It connects quantum error-correction criteria, stabilizer and Calderbank--Shor--Steane codes, and algebraic and homological constructions within a common framework.
Asymptotically good quantum low density parity check codes provide the principal case study.
We formalize constructions and asymptotic parameter theorems spanning hypergraph products, balanced products, lifted products, and quantum Tanner codes.
Further developments cover Bravyi--Poulin--Terhal bounds and related locality restrictions, constant overhead magic state distillation, and energy barriers and storage errors for quantum memories.
QECert establishes a reusable formal foundation for studying the capabilities, resource requirements, and fundamental limitations of quantum error correction, supporting systematic verification and further development of its theory.
\end{abstract}

\section{Introduction}
\label{sec:introduction}

Quantum error correction (QEC) provides a foundation for reliable quantum information processing in the presence of noise.
By encoding information into a larger physical system, a quantum code can protect logical states against errors affecting its constituent parts~\cite{Gott97,Terhal15}.
Recent experiments have demonstrated quantum error-correcting codes and operations on encoded states across several hardware platforms~\cite{LatticeSurgery26,ColorThreshold26,ToricExperiment26,LowOverheadExperiment26,MagicCultivation26,UniversalArchitecture25,SurfaceThreshold25}.
The theory studies how much information can be protected, which errors can be corrected, and what physical resources are required.
The answers determine the resources needed for reliable quantum computation, communication, and storage as physical systems grow.

The mathematical problems arising from these questions range from constructing codes with prescribed properties to determining how noise and locality limit the protection of quantum information.
Modern QEC theory draws on several areas of mathematics and physics, including algebraic topology, algebraic geometry, spectral graph theory, and quantum many-body physics.
For example, graph expansion and local code constraints yield quantum Tanner codes with positive rate and linear distance~\cite{LZ22}.
Geometric embeddings of stabilizer codes and bounds on thermal fluctuations establish the existence of a self-correcting quantum memory in three dimensions~\cite{BDL26}.
Formalizing such arguments allows researchers to verify, step by step, that the conclusion follows from results drawn from different fields.
Lean~\cite{Lean4} provides a language for expressing definitions and proofs, with proof terms checked by its kernel, while Mathlib~\cite{Mathlib20} supplies reusable mathematical results.
Domain libraries such as CSLib~\cite{CSLib26} and Physlib~\cite{Physlib26} extend this approach to computer science and physics.
In quantum computing, formal developments cover programs and circuits~\cite{QWIRE18,Isabelle21,SQIR21,Qbricks20,CoqQ22,QEDIA26}, algorithms~\cite{Shor23,ShorLean26,LeanQuantumAlg26}, and quantum information theory~\cite{QStein25,LeanQuantum26,LeanQIT26}.
These developments also support benchmarks for theorem proving in quantum algorithms and quantum information~\cite{QuantumBench26}.

Within QEC, existing work addresses the verification of correction programs~\cite{QECV21,VeriQEC25}, fault tolerance of logical operations~\cite{FTQEC25}, and stabilizer codes in Coq~\cite{StabilizerCoq25}.
Lean-QEC~\cite{EndToEndQEC26} develops stabilizer-code foundations and a verified reduction to Boolean satisfiability for certifying code distances.
QECLean~\cite{QECLean26} develops a stabilizer and homological framework with verified parameters and logical operators for individual codes and code families.
These projects establish important foundations for code verification.
However, these developments do not yet provide the systematic formal treatment of asymptotic constructions and fundamental limitations needed to verify the theoretical guarantees of scalable QEC.

In this work, we present QECert, a Lean library for the asymptotic properties and fundamental limits of quantum error correction.
We build a common framework for quantum error-correction criteria, stabilizer and Calderbank--Shor--Steane (CSS) codes, and algebraic and homological code constructions.
We validate this framework through formalizations of quantum low density parity check (qLDPC) constructions and their asymptotic properties, with asymptotically good codes~\cite{PK22,LZ22} as the principal case study.
The development also covers Bravyi--Poulin--Terhal (BPT) bounds and related locality restrictions, constant overhead magic state distillation, and energy barriers and storage errors for quantum memories.
Recent advances in AI-assisted mathematics~\cite{AnthropicZeta26,OpenAILocal26,OpenAINS26} suggest that AI could also help researchers tackle open questions in QEC, from the design of quantum codes to the limits of quantum information storage.
By formalizing established QEC theorems in Lean, QECert provides a foundation for deriving and verifying new results in AI-assisted research on quantum error correction.

Section~\ref{sec:architecture} summarizes the main results, compares QECert with existing Lean libraries for QEC, and describes the library architecture.
Section~\ref{sec:foundations} introduces the foundations and code constructions used in the qLDPC development, and Section~\ref{sec:qldpc} presents their asymptotic parameter theorems.
Section~\ref{sec:applications} discusses locality, distillation, and quantum memories.
Section~\ref{sec:conclusion} concludes the paper.
Appendix~\ref{appendix:statements} supplies additional Lean declarations for selected results.
\section{Main Results}
\label{sec:architecture}

QECert brings together formal results on code constructions, locality restrictions, magic state distillation, and quantum memories.
Table~\ref{tab:library-coverage} collects the principal results in the library, with their source references.
The qLDPC development in Section~\ref{sec:qldpc} is the main case study, using the definitions and constructions of Section~\ref{sec:foundations}.
Section~\ref{sec:applications} presents the locality, distillation, and memory results.

\begin{table}[htbp]
\centering
\small
\renewcommand{\arraystretch}{1.12}
\caption{Principal results formalized in QECert.}
\label{tab:library-coverage}
\begin{tabularx}{\linewidth}{@{}p{0.21\linewidth}>{\raggedright\arraybackslash}X@{}}
\toprule
Area & Principal results \\
\midrule
Error correction\newline \S\S~\ref{sec:recovery}--\ref{sec:homology} & Knill--Laflamme theorem~\cite{KL96}; stabilizer codes and CSS codes~\cite{Gott97}. \\
\addlinespace
qLDPC codes\newline \S\S~\ref{sec:products}--\ref{sec:tanner-construction}, \S~\ref{sec:qldpc} & Hypergraph product codes~\cite{TZ09}; quantum expander codes~\cite{LTZ15}; fiber bundle codes~\cite{HHO20}. \\
& Balanced product codes~\cite{BE20}; lifted product codes~\cite{PK20,PK22}; quantum Tanner codes~\cite{LZ22}. \\
& Quantum Tanner codes from lifted products~\cite{LZDec22}. \\
\addlinespace
Locality bounds\newline \S~\ref{sec:locality} & BPT bounds~\cite{BPT10}; separator and connectivity bounds~\cite{BK22,BGKL24}; bounds on long-range interactions~\cite{DL24}. \\
\addlinespace
Distillation and gates\newline \S~\ref{sec:distillation} & Magic state distillation thresholds~\cite{BK05}; triorthogonal codes~\cite{BH12}; sublogarithmic overhead distillation~\cite{HH18}; low-overhead qudit distillation~\cite{KT19}. \\
& Constant overhead distillation~\cite{WHY24}; good binary codes with transversal CCZ gates~\cite{Ngu24}. \\
\addlinespace
Quantum memories\newline \S~\ref{sec:memory} & Bravyi--Terhal energy-barrier bound~\cite{BT09}; energy barriers in the 4D toric code~\cite{DKLP02}; protected observables and storage error bounds~\cite{CLBT10}. \\
\bottomrule
\end{tabularx}
\end{table}

Table~\ref{tab:library-comparison} compares the scope of QECert with two existing Lean libraries for QEC.
The comparison highlights their research focus and representative results, including the code families and reusable foundations developed in earlier work.

\begin{table}[htbp]
\centering
\small
\renewcommand{\arraystretch}{1.12}
\caption{Research focus and representative formal results of three Lean libraries for QEC, based on the cited paper and repository documentation.}
\label{tab:library-comparison}
\begin{tabularx}{\linewidth}{@{}>{\raggedright\arraybackslash}p{0.16\linewidth}>{\raggedright\arraybackslash}p{0.27\linewidth}>{\raggedright\arraybackslash}X@{}}
\toprule
Library & Research focus & Representative results \\
\midrule
Lean-QEC~\cite{EndToEndQEC26} & Distance certification for stabilizer codes & Stabilizer-code foundations; verified reduction to Boolean satisfiability; distance certificates for CSS and bicycle codes. \\
\addlinespace
QECLean~\cite{QECLean26} & Code structures, parameters, and logical operators & Stabilizer and homological frameworks; verified parameters and logical operators for individual codes and families, including toric, surface, and bicycle codes. \\
\addlinespace
QECert (this work) & Asymptotic properties and fundamental limits of QEC & Asymptotically good qLDPC codes; locality restrictions; distillation resource costs; energy barriers and storage error bounds. \\
\bottomrule
\end{tabularx}
\end{table}

\begin{figure}[htbp]
\centering
\includegraphics[width=\linewidth]{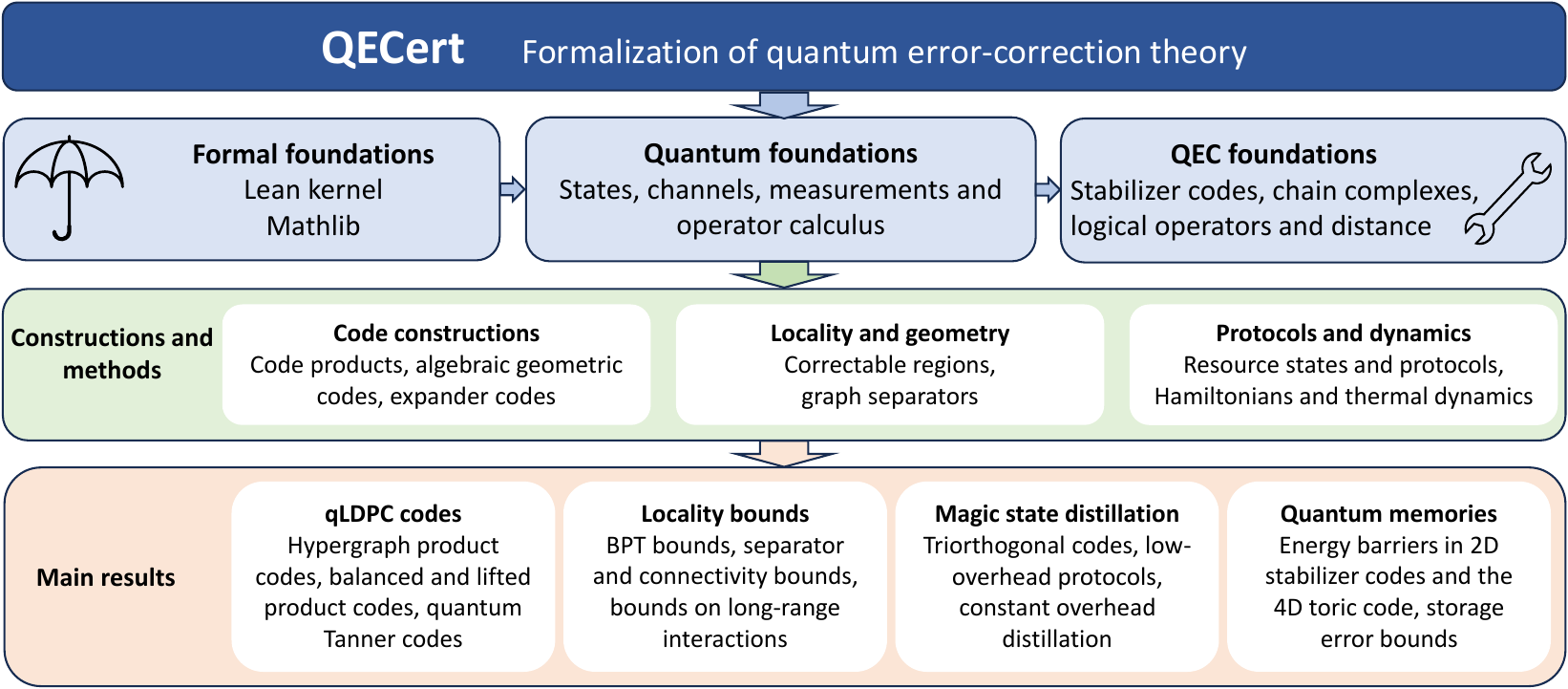}
\caption{Overview of QECert. Shared quantum and coding definitions support code constructions and proof methods, which are used to establish results in four areas of QEC theory. Arrows indicate the main mathematical dependencies.}
\label{fig:architecture}
\end{figure}

The library supports these results through three architectural layers, shown in Figure~\ref{fig:architecture}.
The first layer supplies the quantum and algebraic foundations.
The \lean{QEC.Semantics} modules define states, channels, measurements, and error-correction criteria, while \lean{QEC.Coding} defines classical codes, stabilizer and CSS codes, chain complexes, and distance.
Theorems connect algebraic descriptions of logical operators and distance to error detection on encoded states.
These definitions build on Mathlib, with proofs checked by the Lean kernel.

The second layer develops code constructions and reusable proof methods from these foundations.
Product codes, expander constructions, and algebraic geometric codes come with finite parameter bounds.
Constructions expressed as chain complexes reuse the common CSS commutation and homology results.
The \lean{QEC.Locality} modules supply correctable regions and graph partitions for locality arguments.
The \lean{QEC.Bounds} modules provide the disentangling and partition lemmas used in proofs of geometric tradeoffs.
The \lean{QEC.Distillation} modules specify resource states and protocols, while the dynamics modules in \lean{QEC.Semantics} provide Hamiltonians and thermal evolution.
These models allow coding results to be applied to information processing and storage.

The third layer, \lean{QEC.Results}, groups the principal theorems into the four areas shown in the bottom row of Figure~\ref{fig:architecture}.
In its \lean{QLDPC} module, for example, the balanced product theorem uses finite parameter estimates from \lean{QEC.Coding} to establish the two asymptotic regimes in Theorem~\ref{thm:balanced}.
The \lean{LocalityBounds} module brings together geometric and connectivity bounds.
The \lean{Distillation} module relates code parameters to protocol thresholds and resource costs, while \lean{QuantumMemory} gives energy barrier and storage error bounds.

The library is available at \url{https://github.com/QudeLeap/Lean-QECert}.
The root import \lean{QEC} includes the domain modules and \lean{QEC.Results}.
Each of the four result modules can also be imported separately.
\section{Quantum Codes and Their Constructions}
\label{sec:foundations}

This section introduces the definitions and constructions used in the qLDPC results of Section~\ref{sec:qldpc}.
We begin with quantum error-correction criteria and stabilizer codes, then relate CSS codes to chain complexes and their logical operators to homology.
These common definitions support the product constructions and quantum Tanner codes developed below.

\noindent\textbf{Notation and conventions.}
All Hilbert spaces in this paper have finite dimension.
We write $A^\dagger$ for the adjoint of an operator and $A^T$ for the transpose of a matrix in the chosen bases.
The symbol $F$ denotes a field, with $\mathbb F_2$ used for binary codes.
For a linear map $A$, $\ker A$ and $\operatorname{im}A$ denote its kernel and image.
The Hamming weight $\operatorname{wt}(v)$ of a coordinate vector $v$ counts its nonzero coordinates.
For a Pauli operator, weight counts the physical qubits on which it acts nontrivially, irrespective of its overall phase.
A quantum code with parameters $[[n,k,d]]$ encodes $k$ logical qudits into $n$ physical qudits of a fixed local dimension and has distance $d$; for binary codes these are qubits.

Each Lean listing follows its mathematical definition or theorem.
Proofs and imports are omitted; parameters inherited from the surrounding Lean section are explained in the text.
Longer declarations appear in Appendix~\ref{appendix:statements}.

\subsection{Quantum codes and error correction}
\label{sec:recovery}

Error correction asks whether information can be recovered after a physical noise process.
A recovery operation is a quantum channel, and its existence can be characterized by algebraic conditions on the error operators restricted to the codespace.
The Knill--Laflamme criterion~\cite{KL96} gives these conditions and underlies the notion of detectable errors used below.

A quantum code is a subspace $C\subseteq\mathcal H$ with orthogonal projector $\Pi_C$.
A noise channel $\mathcal N$ is correctable on $C$ if there is a completely positive trace preserving (CPTP) map $\mathcal R$ such that
\begin{equation}
  \mathcal R\circ\mathcal N(\rho)=\rho
  \quad\text{whenever }\rho=\Pi_C\rho\Pi_C.
  \label{eqn:exact-recovery}
\end{equation}

\begin{theorem}[Knill--Laflamme criterion~\cite{KL96}]
\label{thm:knill-laflamme}
Let $\mathcal N(\rho)=\sum_i E_i\rho E_i^\dagger$ with $\sum_i E_i^\dagger E_i=I$.
The channel $\mathcal N$ is correctable on $C$ if and only if there are scalars $\lambda_{ij}$ such that
\begin{equation}
  \Pi_C E_i^\dagger E_j\Pi_C=\lambda_{ij}\Pi_C
  \qquad\text{for all }i,j.
  \label{eqn:knill-laflamme}
\end{equation}
\end{theorem}

Listing~\ref{lst:kl-full} expresses the projector condition using matrix elements between code vectors.
It quantifies over CPTP recovery maps on the same physical Hilbert space $\mathcal H$.
The finite type $\iota$ indexes the Kraus operators, and \lean{E.Complete} expresses trace preservation.

\begin{leanlisting}{Theorem~\ref{thm:knill-laflamme} in Lean}{lst:kl-full}
-- QEC.Semantics.ErrorCorrection.KLRecovery:1654-1665
theorem matrixElementKnillLaflamme_iff_exists_correctingQuantumChannel
    (E : FiniteKrausFamily ι H H) (C : QuantumCode H) (hE : E.Complete) :
    (∃ scalar : ι → ι → Complex,
      ∀ i j : ι,
        ∀ x : H, x ∈ C.carrier →
          ∀ y : H, y ∈ C.carrier →
            inner Complex (E.operator i x) (E.operator j y) =
              scalar i j * inner Complex x y) ↔
      ∃ recovery : QuantumChannel H H,
        ∀ rho : DensityOperator H,
          DensityOperator.SupportedOn C rho →
            recovery (E.inducedMap rho.toOperator) = rho.toOperator
\end{leanlisting}

\subsection{Stabilizer codes and distance}
\label{sec:stabilizer}

For stabilizer codes, this algebraic description becomes particularly concrete.
Pauli operators form a basis for errors, and their commutation relations determine which errors change a syndrome and which preserve the codespace~\cite{Gott97}.
Passing to binary labels retains both this commutation information and the physical support of an operator.

Let $S$ be an abelian subgroup of the Pauli group on $n$ qubits, with $-I\notin S$.
Its common $+1$ eigenspace is a stabilizer code~\cite{Gott97}.
Modulo phases, Pauli operators are vectors $e=(x,z)\in\mathbb F_2^{2n}$, with weight
$\operatorname{wt}(e)=|\{i:(x_i,z_i)\ne(0,0)\}|$.
Commutation is determined by the symplectic form $\langle(x,z),(x',z')\rangle_s=x\cdot z'+z\cdot x'$.
Write $\overline S$ for the binary stabilizer space and $\overline S^{\perp_s}$ for its symplectic orthogonal complement.

\begin{theorem}[Pauli commutation and the symplectic form]
\label{thm:symplectic}
For binary Pauli labels $e,f$, their operator representatives satisfy
\begin{equation}
 E_eE_f=(-1)^{\langle e,f\rangle_s}E_fE_e.
 \label{eqn:symplectic-commutation}
\end{equation}
Thus stabilizer commutation is equivalent to symplectic orthogonality.
\end{theorem}

In Listing~\ref{lst:symplectic}, \lean{Commutes P Q} means that the symplectic pairing vanishes.
The two statements give the commuting and anticommuting cases of Theorem~\ref{thm:symplectic}.

\begin{leanlisting}{Theorem~\ref{thm:symplectic} in Lean}{lst:symplectic}
-- QEC.Coding.Stabilizer.PauliSemantics; proofs omitted
theorem toQubitOperator_commute_of_commutes (P Q : Binary ι) (h : Commutes P Q) :
    toQubitOperator P * toQubitOperator Q = toQubitOperator Q * toQubitOperator P

theorem toQubitOperator_anticommute_of_not_commutes (P Q : Binary ι)
    (h : ¬ Commutes P Q) :
    toQubitOperator P * toQubitOperator Q = -(toQubitOperator Q * toQubitOperator P)
\end{leanlisting}

\begin{definition}[Stabilizer distance]
\label{def:stabilizer-distance}
For a stabilizer code $C$ encoding at least one qubit, its distance is the minimum weight of a Pauli error that cannot be detected,
\begin{equation}
  d(C)=\min\bigl\{\operatorname{wt}(e):
         \Pi_C E_e\Pi_C\notin\mathbb C\Pi_C\bigr\}.
  \label{eqn:stabilizer-distance}
\end{equation}
Here $E_e$ is any Pauli representative of $e$.
\end{definition}

\begin{theorem}[Algebraic characterization of code distance]
\label{thm:stabilizer-distance}
For the stabilizer code of Definition~\ref{def:stabilizer-distance},
\begin{equation}
 d(C)=\min_{e\in\overline S^{\perp_s}\setminus\overline S}\operatorname{wt}(e).
 \label{eqn:algebraic-distance}
\end{equation}
\end{theorem}

Listing~\ref{lst:semantic-distance} equates the physical and algebraic descriptions of distance in Theorem~\ref{thm:stabilizer-distance}.
The predicate on the left states that every Pauli error of smaller weight is detectable and that an undetectable error attains the stated weight.
The right hand side expresses the same minimum over the binary normalizer outside the stabilizer.
The type \lean{SourceValidBinary} includes the conditions ensuring that the stabilizer presentation defines the intended codespace.

\begin{leanlisting}{Theorem~\ref{thm:stabilizer-distance} in Lean}{lst:semantic-distance}
-- QEC.Coding.Stabilizer.Distance:52-62
theorem semanticStabilizerDistance_iff_minWeight_inBinaryNormalizer_diff
    (Q : SourceValidBinary ι) (d : Nat) :
    SemanticStabilizerDistance Q d ↔
      ∃ E : Binary ι,
        Q.1.InBinaryNormalizer E ∧
        ¬ Q.1.InGeneratedGroupUpToSign E ∧
        weight E = d ∧
        ∀ F : Binary ι,
          Q.1.InBinaryNormalizer F →
          ¬ Q.1.InGeneratedGroupUpToSign F →
          d ≤ weight F
\end{leanlisting}

\subsection{CSS codes and homology}
\label{sec:homology}

A Calderbank--Shor--Steane (CSS) code is specified by two families of commuting $X$ and $Z$ checks.
The algebraic constructions below express their commutation condition through a chain complex.
In the $Z$ sector, a cycle represents a Pauli operator with zero syndrome, while a boundary represents a stabilizer.
Two cycles that differ by a boundary have the same logical action.
The homology quotient therefore describes logical $Z$ operators; the dual complex describes the $X$ sector.
The coordinate bases let us measure the weight of each representative and hence the code distance.

\begin{definition}[Chain complex for a CSS code]
\label{def:chain-complex}
A chain complex of based vector spaces over a field $F$ consists of linear maps
\begin{equation}
  C_2\xrightarrow{\partial_2}C_1\xrightarrow{\partial_1}C_0,
  \qquad \partial_1\partial_2=0.
  \label{eqn:chain-complex}
\end{equation}
The basis of $C_1$ labels physical coordinates, and transposes are taken in the chosen bases.
The check matrices are $H_X=\partial_1$ and $H_Z=\partial_2^T$.
Over $\mathbb F_2$, they define the quantum code
\begin{equation}
 Q(H_X,H_Z)=\{\psi:X(h)\psi=\psi\ \forall h\in\operatorname{row}H_X,
 \quad Z(h)\psi=\psi\ \forall h\in\operatorname{row}H_Z\},
 \label{eqn:css-codespace}
\end{equation}
where $X(h)=\bigotimes_i X_i^{h_i}$ and $Z(h)=\bigotimes_i Z_i^{h_i}$ are Pauli operators on the physical qubits.
In our convention, the $X$ and $Z$ check spaces are $\operatorname{im}\partial_1^T$ and $\operatorname{im}\partial_2$, respectively.
\end{definition}

The structure in Listing~\ref{lst:chain-complex} stores the spaces, boundary maps and the condition $\partial_1\partial_2=0$.
It is used over $\mathbb F_2$ for binary CSS codes and over other finite fields for the corresponding qudit constructions.

\begin{leanlisting}{Definition~\ref{def:chain-complex} in Lean}{lst:chain-complex}
-- QEC.Coding.Core.Definitions:596-601
structure ChainComplexData (F : Type u) (C0 C1 C2 : Type x) [Field F]
    [AddCommGroup C0] [Module F C0] [AddCommGroup C1] [Module F C1]
    [AddCommGroup C2] [Module F C2] where
  boundary1 : C1 →ₗ[F] C0
  boundary2 : C2 →ₗ[F] C1
  boundary_squared : boundary1.comp boundary2 = 0
\end{leanlisting}

\begin{theorem}[Logical operators and code distance]
\label{thm:homological-distance}
For the binary CSS code associated with Definition~\ref{def:chain-complex}, the logical $Z$ and $X$ spaces are
\begin{equation}
  \mathcal L_Z\cong H_1=
     \frac{\ker\partial_1}{\operatorname{im}\partial_2},
  \qquad
  \mathcal L_X\cong H^1=
     \frac{\ker\partial_2^T}{\operatorname{im}\partial_1^T}.
  \label{eqn:logical-homology}
\end{equation}
For nonzero logical sectors, their minimum weights are
\begin{align}
 d_Z&=\min\{\operatorname{wt}(z):z\in\ker\partial_1\setminus\operatorname{im}\partial_2\},\\
 d_X&=\min\{\operatorname{wt}(x):x\in\ker\partial_2^T\setminus\operatorname{im}\partial_1^T\}.
\end{align}
The code parameters are
\begin{equation}
  k=\dim_{\mathbb F_2}H_1,\qquad d=\min\{d_X,d_Z\}.
  \label{eqn:homological-distance}
\end{equation}
\end{theorem}

Listing~\ref{lst:homological-distance} relates logical classes to homology while retaining their representatives, whose Hamming weights determine distance.
The hypotheses \lean{hX} and \lean{hZ} supply the two sector minima; estimates for these minima are proved separately for each construction.
The library also identifies the codespace dimension with $2^k$, where $k=\dim_{\mathbb F_2}H_1$.

\begin{leanlisting}{Theorem~\ref{thm:homological-distance} in Lean}{lst:homological-distance}
-- QEC.Coding.Stabilizer.PauliSemantics:2981-2997
theorem binaryCSSCode_homologicalDistance
    {ι κ lambda : Type u} [Fintype ι] [DecidableEq ι] [Fintype κ] [Fintype lambda]
    (K : ChainComplexData
      (ZMod 2) (Word (ZMod 2) κ) (Word (ZMod 2) ι) (Word (ZMod 2) lambda))
    {dX dZ : Nat}
    (hX : PureXLogicalMinimumDistance
      (K.binaryCSSCode).xCode (K.binaryCSSCode).zCode dX)
    (hZ : PureZLogicalMinimumDistance
      (K.binaryCSSCode).xCode (K.binaryCSSCode).zCode dZ) :
    (∀ x : K.middleBoundaryCode.dual.carrier,
      K.pureXLogicalQuotientEquivMiddleCohomology (Submodule.Quotient.mk x) =
        Submodule.Quotient.mk (K.middleBoundaryDualEquivMiddleCocycleCode x)) ∧
    (∀ z : K.firstCoboundaryCheckCode.dual.carrier,
      K.pureZLogicalQuotientEquivMiddleHomology (Submodule.Quotient.mk z) =
        Submodule.Quotient.mk (K.firstCoboundaryDualEquivMiddleCycleCode z)) ∧
    AlgebraicCSSMinimumDistance
      (K.binaryCSSCode).xCode (K.binaryCSSCode).zCode (min dX dZ)
\end{leanlisting}

\subsection{Hypergraph and lifted products}
\label{sec:products}

A classical linear code with parity check matrix $H\in F^{m\times n}$ is $\ker H$; its dimension is $\dim\ker H$ and its distance is the smallest weight of a nonzero codeword.
Its dual is taken with respect to the coordinate pairing.
Sparse parity checks provide the input to the following product constructions.
The difficulty is that two independently chosen sparse matrices need not define commuting quantum checks.
The hypergraph product~\cite{TZ09} resolves this by combining two classical matrices in a tensor product complex.
The two contributions to the composition of boundary maps cancel, which ensures that the quantum checks commute.
Sparsity of the input matrices controls the weights of those checks.

\begin{definition}[Hypergraph product~\cite{TZ09}]
\label{def:hypergraph-product}
For binary matrices $H_i\in\mathbb F_2^{m_i\times n_i}$, define
\begin{equation}
 H_X=\begin{bmatrix}H_1\otimes I_{n_2}&I_{m_1}\otimes H_2^T\end{bmatrix},\qquad
 H_Z=\begin{bmatrix}I_{n_1}\otimes H_2&H_1^T\otimes I_{m_2}\end{bmatrix}.
 \label{eqn:hgp-checks}
\end{equation}
These matrices satisfy $H_XH_Z^T=0$.
The hypergraph product code is $Q(H_X,H_Z)$ on $n_1n_2+m_1m_2$ coordinates, with $Q$ defined in Equation~\eqref{eqn:css-codespace}.
\end{definition}

The associated complex is the tensor product of the two short complexes with boundaries $H_1$ and $H_2^T$.
This interpretation extends to matrices over a group algebra~\cite{PK20,PK22}.
An entry of such a matrix specifies a linear combination of translations on a group coordinate.
Replacing entries by their regular actions produces ordinary matrices over $F$, so the resulting code still has a concrete finite set of physical coordinates.
The two factors share a group coordinate, so increasing the group size increases the block length.
The supports of the group algebra entries, together with the matrix sparsity, control the check weights.

\begin{definition}[Lifted product]
\label{def:lifted-data}
Let $F$ be a finite field, let $G$ be a finite group, and let $A\in F[G]^{m_A\times n_A}$ and $B\in F[G]^{m_B\times n_B}$.
For $a=\sum_{g\in G}a_g g\in F[G]$, define regular matrices by
$(R_a)_{u,v}=a_{v^{-1}u}$ and $(L_a)_{u,v}=a_{uv^{-1}}$ for $u,v\in G$.
The block matrices $R_A$ and $L_B$ apply these operations to each entry of $A$ and $B$.
The two boundary maps have blocks
\begin{equation}
 \partial_1=\begin{bmatrix}R_A\otimes I_{m_B}&-I_{m_A}\otimes L_B\end{bmatrix},\qquad
 \partial_2=\begin{bmatrix}I_{n_A}\otimes L_B\\R_A\otimes I_{n_B}\end{bmatrix}.
 \label{eqn:lifted-boundaries}
\end{equation}
Here the tensor notation acts on the row and column indices; every block acts on the same group coordinate.
The middle space has dimension
\begin{equation}
 n=|G|(n_A m_B+m_A n_B).
 \label{eqn:lifted-length}
\end{equation}
\end{definition}

\begin{theorem}[Product boundaries define a chain complex]
\label{thm:product-complex}
The boundaries in Definition~\ref{def:lifted-data} satisfy $\partial_1\partial_2=0$.
They define the lifted product CSS code with checks $H_X=\partial_1$ and $H_Z=\partial_2^T$, using the finite field qudit construction when $F\ne\mathbb F_2$.
\end{theorem}

Left and right regular actions commute, so the two terms in the product cancel, including when $G$ is noncommutative.
For the trivial group, taking $A=H_1$ and $B=H_2^T$ recovers Definition~\ref{def:hypergraph-product} over $\mathbb F_2$.
Listing~\ref{lst:product-complex} gives the matrix inputs and verifies this cancellation for the actual boundary matrices.

\begin{leanlisting}{Definition~\ref{def:lifted-data} and Theorem~\ref{thm:product-complex} in Lean}{lst:product-complex}
-- QEC.Coding.ExpanderCodes.LiftedProduct; proof omitted
structure Data (F G aRow aCol bRow bCol : Type*) [Semiring F] where
  A : Matrix aRow aCol (MonoidAlgebra F G)
  B : Matrix bRow bCol (MonoidAlgebra F G)

theorem boundary1Matrix_mul_boundary2Matrix
    {F G aRow aCol bRow bCol : Type*}
    [Field F] [Fintype F] [Group G] [Fintype G] [DecidableEq G]
    [Fintype aRow] [DecidableEq aRow]
    [Fintype aCol] [DecidableEq aCol]
    [Fintype bRow] [DecidableEq bRow]
    [Fintype bCol] [DecidableEq bCol]
    (D : Data F G aRow aCol bRow bCol) :
    boundary1Matrix D * boundary2Matrix D = 0
\end{leanlisting}

For LDPC constructions we track both the maximum check weight and the maximum number of selected checks containing a coordinate.
The latter is the qubit incidence.
The product locality estimates bound both quantities in terms of the sparsity of the input matrices.

\subsection{Balanced products and subsystem codes}
\label{sec:balanced-construction}

Balanced products~\cite{BE20} use compatible group actions to identify coordinates in a product construction.
The quotient retains the local structure needed for sparse checks while reducing the number of physical coordinates.
This operation can be described either by relations on vector spaces or by a quotient of a product of graphs.

Let $U$ be a right $F[G]$ module and $V$ a left $F[G]$ module.
Their balanced tensor product imposes the following relations for all $u\in U$, $v\in V$ and $g\in G$:
\begin{equation}
 (ug)\otimes v=u\otimes(gv),\qquad
 U\otimes_{F[G]}V=(U\otimes_F V)/\langle ug\otimes v-u\otimes gv\rangle.
 \label{eqn:balanced-tensor}
\end{equation}
The denominator denotes the $F$ linear span of these relations.
Geometrically, graphs $X$ and $Y$ with right and left $G$ actions give a quotient of their product complex,
\begin{equation}
 X\times_G Y=(X\times Y)/{\sim},\qquad (xg,y)\sim(x,gy),
 \label{eqn:balanced-geometric}
\end{equation}
where $x\in X$, $y\in Y$ and $g\in G$; the actions identify vertices, edges and squares compatibly.
The corresponding operation on equivariant chain complexes is the balanced tensor product~\cite{BE20}.

For the asymptotic parameter bounds in Section~\ref{sec:balanced-almost}, we formalize the cyclic specialization using lifted product matrices.
The general balanced product construction is not formalized here.

\begin{definition}[Cyclic balanced product]
\label{def:cyclic-product}
Let $q>1$ be the cycle length and $G=\mathbb Z/q\mathbb Z$, written multiplicatively with generator $x$.
Let $A\in\mathbb F_2[G]^{m_A\times n_A}$ represent the equivariant binary Tanner boundary.
The cyclic construction uses Definition~\ref{def:lifted-data} with
\begin{equation}
 B=[1+x],\qquad n=q(n_A+m_A).
 \label{eqn:cyclic-product}
\end{equation}
\end{definition}

Listing~\ref{lst:cyclic-product} constructs the corresponding lifted product data with $B=[1+x]$, reusing the boundary maps defined above.

\begin{leanlisting}{Definition~\ref{def:cyclic-product} in Lean}{lst:cyclic-product}
-- QEC.Coding.BalancedProduct.Cyclic
noncomputable def cyclicBoundaryMatrix {G : Type u} [Group G] (x : G) :
    Matrix Unit Unit (MonoidAlgebra (ZMod 2) G) :=
  fun _ _ => cyclicBoundaryElement x

noncomputable def liftedProductData
    {G : Type u} {aRow aCol : Type v} [Group G]
    (A : Matrix aRow aCol (MonoidAlgebra (ZMod 2) G)) (x : G) :
    LiftedProduct.Data (ZMod 2) G aRow aCol Unit Unit where
  A := A
  B := cyclicBoundaryMatrix x
\end{leanlisting}

The cyclic complex first defines a CSS code by Equation~\eqref{eqn:css-codespace}.
To obtain the subsystem code used in Section~\ref{sec:balanced-almost}, the construction then selects the horizontal logical sectors as protected information and assigns the vertical sectors to gauge degrees of freedom~\cite{BE20}.
For a subsystem code $C\cong\mathcal H_L\otimes\mathcal H_G$, $\mathcal H_L$ carries the protected information and $\mathcal H_G$ the gauge subsystem.
A dressed logical operator acts nontrivially on $\mathcal H_L$ and may also act on $\mathcal H_G$; its distance is minimized over these gauge choices.
The distances in the balanced product theorem refer to these dressed logical operators.

\subsection{Cayley complexes and quantum Tanner codes}
\label{sec:tanner-construction}

Quantum Tanner codes~\cite{LZ22} use local classical codes on a complex built from left and right Cayley graphs.
The classical Tanner construction places a small code around each vertex and requires overlapping local words to agree.
The quantum construction must also ensure that the two types of checks commute.
The square complex supplies a common set of qubits and a controlled pattern of overlaps, while dual local codes enforce the required orthogonality.
We first define the checks; Section~\ref{sec:tanner} gives the expansion and robustness conditions needed for the distance bound.

For a finite group $G$ and symmetric generating sets $A,B\subset G$ excluding the identity, the edges connect $g$ to $ag$ and to $gb$.
We use the nondegeneracy condition $ag\ne gb$ for every $a\in A$, $b\in B$ and $g\in G$.
The commuting left and right actions give squares with corners $g,ag,gb,agb$.
The construction uses copies of the vertex set to distinguish the two types of checks and specifies which squares are incident on each vertex.
Write $\mathcal Q$ for the square set and $V_X,V_Z$ for the two vertex types.
The physical qubits are indexed by $\mathcal Q$, and the squares incident on each vertex are indexed by $A\times B$.

\begin{definition}[Local tensor code]
\label{def:local-tensor}
Let $C_A\subseteq\mathbb F_2^A$ and $C_B\subseteq\mathbb F_2^B$ be classical codes.
Their tensor code is
\begin{equation}
 C_A\otimes C_B=\{c\in\mathbb F_2^{A\times B}:
 c(\cdot,b)\in C_A\ \forall b\in B,\quad c(a,\cdot)\in C_B\ \forall a\in A\}.
 \label{eqn:tensor-code}
\end{equation}
\end{definition}

\begin{definition}[Quantum Tanner code]
\label{def:tanner-checks}
For $v\in V_X\cup V_Z$, let $\iota_v:\mathbb F_2^{A\times B}\to\mathbb F_2^{\mathcal Q}$ extend a local word to the incident squares and set it to zero elsewhere.
The global check spaces are
\begin{equation}
 \mathcal S_X=\sum_{v\in V_X}\iota_v(C_A^\perp\otimes C_B^\perp),\qquad
 \mathcal S_Z=\sum_{v\in V_Z}\iota_v(C_A\otimes C_B).
 \label{eqn:tanner-checks}
\end{equation}
The square incidence and local duality ensure $\mathcal S_X\perp\mathcal S_Z$~\cite{LZ22}.
The quantum Tanner code is their common Pauli $+1$ eigenspace, as in Equation~\eqref{eqn:css-codespace}.
\end{definition}

Listing~\ref{lst:tanner-construction} shows how a basis of each local tensor code becomes a finite family of physical checks.
The structure \lean{RawData} records the local tensor code bases explicitly.
The Boolean coordinate selects $V_X$ or $V_Z$, and \lean{Square} is the physical coordinate set $\mathcal Q$.
The final definition uses \lean{Data}, which adds the proved boundary commutation law to \lean{RawData}.
The resulting boundary maps give the CSS complex of Section~\ref{sec:homology}, so its logical dimension and distances have the definitions already given there.

\begin{leanlisting}{From local tensor bases to the quantum Tanner code of Definition~\ref{def:tanner-checks}}{lst:tanner-construction}
-- QEC.Coding.QuantumTanner.Core
structure RawData
    (G A B Square XBasis ZBasis : Type v)
    [Group G] [Fintype G] [Fintype A] [Fintype B] [Fintype Square]
    [Fintype XBasis] [Fintype ZBasis] where
  complex : LeftRightCayleySquareComplex G A B Square
  codeA : LinearCode (ZMod 2) A
  codeB : LinearCode (ZMod 2) B
  xBasis : Module.Basis XBasis (ZMod 2) (tensorCode codeA.dual codeB.dual).carrier
  zBasis : Module.Basis ZBasis (ZMod 2) (tensorCode codeA codeB).carrier

namespace RawData
def xGenerator (D : RawData G A B Square XBasis ZBasis) (check : G × XBasis) :
    Word (ZMod 2) Square :=
  D.complex.localExtension (check.1, true) (D.xBasis check.2).1

def zGenerator (D : RawData G A B Square XBasis ZBasis) (check : G × ZBasis) :
    Word (ZMod 2) Square :=
  D.complex.localExtension (check.1, false) (D.zBasis check.2).1

end RawData

namespace Data

def cssCode (D : Data G A B Square XBasis ZBasis) :=
  D.chainComplex.binaryCSSCode

end Data
\end{leanlisting}
\section{Toward Asymptotically Good Quantum LDPC Codes}
\label{sec:qldpc}

A central goal in quantum coding theory is to construct codes that combine positive rate, linear distance and bounded check weights.
The constructions in Section~\ref{sec:foundations} approach this goal through different tradeoffs between dimension and distance.
We now state their parameter guarantees and present the corresponding Lean theorems.

The hypergraph product construction~\cite{TZ09} gives positive rate and distance of order $\sqrt n$ from suitable classical LDPC codes.
Balanced product codes~\cite{BE20} and almost linear lifted product codes~\cite{PK20} improve the attainable relations between dimension and distance.
Good lifted product codes~\cite{PK22} and quantum Tanner codes~\cite{LZ22} reach the target $k=\Omega(n)$ and $d=\Omega(n)$.
Table~\ref{tab:qldpc-results} collects the parameter regimes considered below.

We use $O$, $\Omega$ and $\Theta$ for asymptotic bounds as the physical length $n$ tends to infinity.
Their constants may depend on fixed local codes, graph degrees and target rates, but are independent of the growing construction index.

Let $\{Q_a\}$ be a family with parameters $[[n_a,k_a,d_a]]$ over a fixed local dimension and with unbounded physical length.
Here $k_a$ counts encoded qudits, or logical qubits in the binary case.
For a chosen check presentation, let $w_a$ be the largest check weight and $v_a$ the largest number of checks incident on a physical coordinate.

\begin{definition}[Asymptotically good qLDPC family]
\label{def:good-family}
The family is qLDPC if there is a constant $w$ such that $w_a,v_a\le w$ for every $a$.
It is asymptotically good if, in addition, there are constants $R,\delta>0$ and $N$ such that
\begin{equation}
  n_a\ge N\quad\Longrightarrow\quad
  k_a\ge Rn_a\quad\text{and}\quad d_a\ge\delta n_a.
  \label{eqn:good-family}
\end{equation}
\end{definition}

Listing~\ref{lst:good-family} first defines the code family and then its locality, rate and distance predicates.
The type \lean{A} indexes the family; $\iota(a)$ and $\kappa(a)$ index the physical coordinates and checks of member $a$.
The field \lean{cofinal_blockLength} requires arbitrarily large physical lengths.
The rate and distance predicates may use different length thresholds; taking their maximum gives the common threshold in Definition~\ref{def:good-family}.
All constants are fixed before the family index varies.

\begin{leanlisting}{Definition~\ref{def:good-family} in Lean}{lst:good-family}
-- QEC.Coding.QLDPC.QuantumCodeFamily:19-24
structure QuantumCodeFamily
    (A : Type u) (ι : A → Type v) (κ : A → Type w)
    [∀ a, Fintype (ι a)] [∀ a, DecidableEq (ι a)] [∀ a, Fintype (κ a)] where
  presentation : ∀ a, FiniteQuditCheckPresentation (ι a) (κ a)
  cofinal_blockLength : ∀ N : Nat, ∃ a : A, N ≤ (presentation a).blockLength

-- QEC.Coding.QLDPC.AsymptoticallyGoodFamily:22-23
def HasUniformLDPC (F : QuantumCodeFamily A ι κ) : Prop :=
  ∃ B : Nat, ∀ a, Coding.BoundedLocality (F.localityProfile a) B

-- QEC.Coding.QLDPC.AsymptoticallyGoodFamily:26-28
def HasEventualRateAtLeast (F : QuantumCodeFamily A ι κ) (rate : ℝ) : Prop :=
  ∃ N : Nat, ∀ a, N ≤ F.blockLength a →
    rate * (F.blockLength a : ℝ) ≤ (F.encodedDimension a : ℝ)

-- QEC.Coding.QLDPC.AsymptoticallyGoodFamily:31-34
def HasEventualRelativeDistanceAtLeast
    (F : QuantumCodeFamily A ι κ) (relativeDistance : ℝ) : Prop :=
  ∃ N : Nat, ∀ a, N ≤ F.blockLength a →
    relativeDistance * (F.blockLength a : ℝ) ≤ (F.distance a : ℝ)

-- QEC.Coding.QLDPC.AsymptoticallyGoodFamily:38-42
def IsAsymptoticallyGoodQLDPCFamily (F : QuantumCodeFamily A ι κ) : Prop :=
  HasUniformLDPC F ∧
    (∃ rate : ℝ, 0 < rate ∧ HasEventualRateAtLeast F rate) ∧
    (∃ relativeDistance : ℝ, 0 < relativeDistance ∧
      HasEventualRelativeDistanceAtLeast F relativeDistance)
\end{leanlisting}

\begin{table}[H]
\centering
\small
\caption{Representative qLDPC constructions. All bounds refer to physical block length $n$ and hold for sufficiently large members of each family. Balanced product distances are dressed subsystem distances.}
\label{tab:qldpc-results}
\begin{tabularx}{\linewidth}{@{}p{0.29\linewidth}X@{}}
\toprule
Construction & Parameters \\
\midrule
Hypergraph product~\cite{TZ09} & $k=\Theta(n)$, $d=\Theta(\sqrt n)$ \\
Balanced product~\cite{BE20} & $k=\Theta(n^{2/3})$, $d_X=\Omega(n^{1/3})$, $d_Z=\Theta(n)$; also $k=\Theta(n^{4/5})$, $d=\Omega(n^{3/5})$ \\
Almost linear lifted product~\cite{PK20} & $k=\Theta(\log n)$, $d=\Theta(n/\log n)$ \\
Good lifted product~\cite{PK22} & $k\ge Rn$, $d=\Theta(n)$ for any fixed $0<R<1$ \\
Quantum Tanner~\cite{LZ22} & $k=\Omega(n)$, $d=\Omega(n)$ \\
\bottomrule
\end{tabularx}
\end{table}

\subsection{Classical expander codes and expansion properties}
\label{sec:expansion}

Classical expander codes~\cite{SS96} combine a sparse graph with local code constraints to obtain a global code with positive rate and linear distance.
With suitable local codes and sufficiently strong graph expansion, this construction supplies classical inputs for quantum product codes.

\begin{definition}[Classical Tanner code]
\label{def:classical-tanner}
Let $G=(V,E)$ be a finite simple $\Delta$ regular graph with $\Delta>0$, and let $C_0\subseteq F^\Delta$ be a linear code over a finite field $F$.
For each vertex $v$, fix an ordering of the incident edge set $E(v)$, identifying it with the coordinates of $C_0$.
The associated Tanner code is
\begin{equation}
 C(G,C_0)=\{c\in F^E:c|_{E(v)}\in C_0\text{ for every }v\in V\}.
 \label{eqn:classical-tanner}
\end{equation}
\end{definition}

The coordinates of this classical code are the edges, and each vertex imposes the parity checks of $C_0$ on its incident coordinates.
Fixed graph degree gives bounded check weights and incidences.
On an expanding graph, suitable local rate and distance yield an expander code~\cite{SS96,BE20}.
The spectral estimates below explain how the graph contributes to its distance.

Several constructions use the Ramanujan graphs of Lubotzky, Phillips and Sarnak (LPS)~\cite{LPS88}.
For a connected graph of degree $\Delta$ with adjacency matrix $A$, the Ramanujan property is the spectral bound
\begin{equation}
  |\lambda|\le 2\sqrt{\Delta-1}
  \quad\text{for every }\lambda\in\operatorname{spec}(A)\setminus\{\pm\Delta\}.
  \label{eqn:lps-spectrum}
\end{equation}
Here $\operatorname{spec}(A)$ is the adjacency spectrum.
The eigenvalue $\Delta$ corresponds to constant functions; $-\Delta$ occurs when the graph is bipartite.
For the nonbipartite graphs used in the good lifted product construction, the equivalent operator bound is $\|A|_{\mathbf 1^\perp}\|\le2\sqrt{\Delta-1}$.
Here $\mathbf 1^\perp$ is the space of mean zero vertex functions, and the norm is the Euclidean operator norm.
The spectral estimates for the chosen LPS graphs are taken as assumptions in the Lean theorems below.

For example, let $V$ be the vertex set of a nonempty $\Delta$ regular graph and let $\Lambda$ bound its second largest adjacency eigenvalue.
For a vertex subset $U\subseteq V$,
\begin{equation}
 2|E(U,U)|\le\Lambda |U|+
 (\Delta-\Lambda)\frac{|U|^2}{|V|}.
 \label{eqn:spectral-small-set}
\end{equation}
Here $E(U,U)$ denotes the edges with both ends in $U$.
This follows by separating the indicator of $U$ into its constant and mean zero parts.
For the Tanner code $C(G,C_0)$, Equation~\eqref{eqn:spectral-small-set} combines with the local code distance to bound the support of every nonzero codeword.
Suitable local code parameters and graph expansion then give a distance proportional to $|E|$~\cite{BE20}.

The quantum constructions use both these classical codes and stronger properties of their check maps.
For balanced products, the formal development derives small support expansion estimates for the Tanner boundary and its transpose from graph expansion and local code properties.
The almost linear lifted product theorem takes the required matrix expansion as an input~\cite{PK20}.
Good lifted product codes additionally use local product expansion~\cite{PK22}, while quantum Tanner codes use the robustness of local tensor codes described in Section~\ref{sec:tanner}~\cite{LZ22}.

\subsection{Hypergraph product codes}
\label{sec:hypergraph}

The hypergraph product~\cite{TZ09} gives a general way to obtain quantum LDPC codes from classical LDPC codes with positive rate and linear distance.
Its parameter formulas explain the square root distance scale.
For two comparable input lengths, the number of physical qubits grows quadratically, whereas the distance scale inherited from the inputs grows linearly.
This yields the square root distance regime in Table~\ref{tab:qldpc-results} and provides a useful reference for the later constructions.

Let $H_i\in\mathbb F_2^{m_i\times n_i}$, and write $k_i=\dim\ker H_i$ and $k_i^T=\dim\ker H_i^T$.
Let $d_i$ and $d_i^T$ be the corresponding classical distances, with distance $\infty$ for the zero code.
Definition~\ref{def:hypergraph-product} gives the checks; the following theorem relates their parameters to the two classical input codes.

\begin{theorem}[Parameters of hypergraph product codes]
\label{thm:hypergraph}
The code associated with $H_1,H_2$ has
\begin{equation}
  n=n_1n_2+m_1m_2,\qquad
  k=k_1k_2+k_1^Tk_2^T,
  \label{eqn:hgp-parameters}
\end{equation}
and its distance satisfies
\begin{equation}
  d\ge\min\{d_1,d_2,d_1^T,d_2^T\}.
  \label{eqn:hgp-distance}
\end{equation}
If every row and column of both input matrices has weight at most $w$, then every output check has weight at most $2w$, and every qubit belongs to at most $2w$ checks.
\end{theorem}

Listing~\ref{lst:hgp} formalizes Theorem~\ref{thm:hypergraph} for the two input matrices.
It also gives exact formulas for the check weights and incidences.
The extra hypothesis \lean{DistanceEqualityCondition} identifies when the lower bound in Equation~\eqref{eqn:hgp-distance} is attained.

\begin{leanlisting}{Theorem~\ref{thm:hypergraph} in Lean}{lst:hgp}
-- QEC.Results.QLDPC.HypergraphProductParameters:21-60
theorem tillichZemorHypergraphProductParameters
    {rowOne colOne rowTwo colTwo : Type u}
    [Fintype rowOne] [DecidableEq rowOne]
    [Fintype colOne] [DecidableEq colOne]
    [Fintype rowTwo] [DecidableEq rowTwo]
    [Fintype colTwo] [DecidableEq colTwo]
    (D : Data (ZMod 2) rowOne colOne rowTwo colTwo) :
    (binaryCSSCode D).stabilizer.IsSourceValidBinary ∧
    Fintype.card (QubitIndex rowOne colOne rowTwo colTwo) =
      Fintype.card colOne * Fintype.card colTwo +
        Fintype.card rowOne * Fintype.card rowTwo ∧
    Module.finrank (ZMod 2) (chainComplex D).middleHomology =
      Dimension (firstCode D) * Dimension (secondCode D) +
        Dimension (firstTransposeCode D) *
          Dimension (secondTransposeCode D) ∧
    inputDistanceMinimum D ≤
      extendedAlgebraicCSSMinimumDistance
        (binaryCSSCode D).xCode (binaryCSSCode D).zCode ∧
    (DistanceEqualityCondition D →
      extendedAlgebraicCSSMinimumDistance
          (binaryCSSCode D).xCode (binaryCSSCode D).zCode =
        inputDistanceMinimum D) ∧
    (∀ r₁ c₂, HammingWeight
        (wordLinearMapTranspose (boundary1 D) (Pi.single (r₁, c₂) 1)) =
      rowWeight D.first r₁ + columnWeight D.second c₂) ∧
    (∀ c₁ r₂, HammingWeight (boundary2 D (Pi.single (c₁, r₂) 1)) =
      rowWeight D.second r₂ + columnWeight D.first c₁) ∧
    (∀ c₁ c₂, qubitDegree D (Sum.inl (c₁, c₂)) =
      columnWeight D.first c₁ + columnWeight D.second c₂) ∧
    (∀ r₁ r₂, qubitDegree D (Sum.inr (r₁, r₂)) =
      rowWeight D.second r₂ + rowWeight D.first r₁) ∧
    (∀ w,
      (∀ r, rowWeight D.first r ≤ w) →
      (∀ c, columnWeight D.first c ≤ w) →
      (∀ r, rowWeight D.second r ≤ w) →
      (∀ c, columnWeight D.second c ≤ w) →
      ((∀ x₀, HammingWeight
          (wordLinearMapTranspose (boundary1 D) (Pi.single x₀ 1)) ≤ 2 * w) ∧
       (∀ x₂, HammingWeight (boundary2 D (Pi.single x₂ 1)) ≤ 2 * w) ∧
       (∀ q, qubitDegree D q ≤ 2 * w)))
\end{leanlisting}

The two terms in the dimension formula come from the two tensor contributions to the homology of the product complex.
The distance argument relates a nontrivial logical class to a nonzero word in one of the input codes or transpose codes.
The weight of that word is bounded below by the corresponding classical distance~\cite{TZ09}.
Sparsity is checked directly on the two blocks of each check matrix.

When $H_1=H_2$ has full row rank, Equation~\eqref{eqn:hgp-parameters} reduces to
$n=n_1^2+(n_1-k_1)^2$ and $k=k_1^2$.
Classical LDPC families with positive rate and linear distance therefore yield the first row of Table~\ref{tab:qldpc-results}~\cite{TZ09}.

\subsection{Balanced product and almost linear lifted product codes}
\label{sec:balanced-almost}

\paragraph{Balanced product codes.}
Balanced product codes~\cite{BE20} use group symmetry to change the relation between the size of a product construction and the weights of its logical operators.
The cyclic construction of Section~\ref{sec:balanced-construction} combines an equivariant Tanner code with a cycle and then selects a protected subsystem.
The two dressed logical distances scale differently, so we state the $X$ and $Z$ guarantees separately.
The second parameter regime below gives a distance exponent above one half while retaining bounded check weights and incidences.

For the construction in Definition~\ref{def:cyclic-product}, the cyclic matrix construction gives the physical length, while graph expansion and the local classical distance control the logical weights~\cite{BE20}.
These estimates yield the following parameter regimes.

\begin{theorem}[Parameters of balanced product subsystem codes~\cite{BE20}]
\label{thm:balanced}
Using the LPS spectral estimate for the selected graph family, the construction gives qLDPC subsystem codes with asymmetric dressed distances and parameters
\begin{equation}
 k=\Theta(n^{2/3}),\qquad d_X=\Omega(n^{1/3}),\qquad d_Z=\Theta(n).
 \label{eqn:balanced-unbalanced}
\end{equation}
A second construction with balanced distances gives qLDPC subsystem codes with
\begin{equation}
 k=\Theta(n^{4/5}),\qquad d=\Omega(n^{3/5}).
 \label{eqn:balanced-family}
\end{equation}
The length $n$ is unbounded, and the locality bounds are uniform within each family.
\end{theorem}

Listing~\ref{lst:balanced} is the Lean form of Theorem~\ref{thm:balanced}.
The Lean construction fixes the graph degree at $\Delta=402$.
An admissible prime determines the graph, subgroup, local code and finite code member.
For every proposed bound on the prime, the theorem supplies a larger admissible prime and a code member satisfying the parameter estimates.

\begin{leanlisting}{Theorem~\ref{thm:balanced} in Lean}{lst:balanced}
-- QEC.Results.QLDPC.BalancedProductFamily:16-24
theorem breuckmannEberhardtBalancedProductQLDPCFamilies
    (hLPS : LPS401FamilyRayleighBound) :
    (∀ B : Nat, ∃ a : AdmissiblePrime401,
      max B BE20Unbalanced.familyThreshold < a.q ∧
      BE20Unbalanced.UnbalancedFamilyCertificate
        a.q a.gt40 a.legendre) ∧
    (∀ B : Nat, ∃ a : AdmissiblePrime401,
      max B BE20Unbalanced.familyThreshold < a.q ∧
      BalancedFamilyCertificate a.q a.gt40 a.legendre)
\end{leanlisting}

The predicates \lean{UnbalancedFamilyCertificate} and \lean{BalancedFamilyCertificate} express finite inequalities from which the exponents follow.
They include the exact physical length, upper and lower bounds on $k$, lower bounds on dressed distances, and bounded check weights and incidences.
Listing~\ref{lst:be-unbalanced-certificate} and Listing~\ref{lst:be-certificate} give the two certificates and their finite parameter bounds in Appendix~\ref{appendix:balanced-statements}.
These inequalities express the estimates in terms of physical block length, giving the exponents in Theorem~\ref{thm:balanced}.

The distance proof uses the local classical code to constrain a logical representative near each vertex and graph expansion to control how those local constraints can be distributed~\cite{BE20}.
The cyclic direction contributes a different scale, which is reflected in the unequal $X$ and $Z$ distance estimates.
Because vertical logical sectors are assigned to gauge degrees of freedom, the relevant minimization allows changes in those sectors.

\paragraph{Almost linear lifted product codes.}
The almost linear construction~\cite{PK20} attains distance within a logarithmic factor of the physical length, with a logarithmic number of logical qubits.
The more general tradeoff below shows how increasing the encoded dimension changes the available distance guarantee.

It pairs the cyclic factor $1+x$ with a sparse expanding matrix $A$ over a cyclic group algebra~\cite{PK20}.
The star transpose is $A^*=(\overline{A_{ji}})_{i,j}$, where $\overline{\sum_g a_g g}=\sum_g a_g g^{-1}$.
Expansion of $A$ and $A^*$ controls the two logical sectors.
Here expansion requires the image of a vector with small support to have weight at least a fixed multiple of its input weight.
The two matrix estimates enter the lower bounds on $X$ and $Z$ logical weights.
Proposition 1 of Ref.~\cite{PK20} provides the required expanding matrix family.

\begin{theorem}[Quantum LDPC codes with almost linear distance~\cite{PK20}]
\label{thm:almost}
Given the expanding matrix family of Proposition 1, there is a binary qLDPC family with
\begin{equation}
  k=\Theta(\log n),\qquad
  d=\Theta\!\left(\frac{n}{\log n}\right).
  \label{eqn:almost-linear}
\end{equation}
More generally, for every fixed $0\le\alpha<1$, there is a binary qLDPC family with
\begin{equation}
  k=\Omega(n^\alpha\log n),\qquad
  d=\Omega\!\left(\frac{n^{1-\alpha/2}}{\log n}\right).
  \label{eqn:almost-tradeoff}
\end{equation}
\end{theorem}

Listing~\ref{lst:almost} takes the expanding matrix family as its input.
Its first conclusion supplies arbitrarily large cyclic lifts with the parameters in Equation~\eqref{eqn:almost-linear}.
The second gives the family for a chosen exponent $\alpha$.

\begin{leanlisting}{Theorem~\ref{thm:almost} in Lean}{lst:almost}
-- QEC.Coding.ExpanderCodes.AlmostLinear:937-952
def FirstAlmostLinearFamily : Prop :=
  ∃ (w gamma radiusNumerator radiusDenominator
      expansionNumerator expansionDenominator : Nat),
    4 ≤ w ∧
    0 < gamma ∧
    0 < radiusNumerator ∧
    0 < radiusDenominator ∧
    0 < expansionNumerator ∧
    0 < expansionDenominator ∧
    ∀ B : Nat,
      ∃ (q n m : Nat) (_hq : NeZero q)
        (A : Matrix (Fin m) (Fin (w * n))
          (MonoidAlgebra (ZMod 2) (Cyclic.CyclicGroup q))),
        B < q ∧
        FirstFamilyMember w gamma radiusNumerator radiusDenominator
          expansionNumerator expansionDenominator q n m A

-- QEC.Results.QLDPC.LiftedProductAlmostLinear:26-27
def LiftedProductAlmostLinearDistance : Prop :=
  FirstAlmostLinearFamily ∧ TwoProductConstructionFamily

-- QEC.Results.QLDPC.LiftedProductAlmostLinear:30-32
theorem liftedProductAlmostLinearDistance
    (hPropositionOne : PropositionOneExpandingMatrixFamily) :
    LiftedProductAlmostLinearDistance
\end{leanlisting}

The finite estimates in Ref.~\cite{PK20} control both CSS distance sectors and include an explicit logical operator giving the upper distance bound in Equation~\eqref{eqn:almost-linear}.
If $\ell$ is the cyclic lift size, the selected base dimensions grow as $\Theta(\log\ell)$.
Together, the length, dimension and distance estimates give
\begin{equation}
 n=\Theta(\ell\log\ell),\qquad k=\Theta(\log\ell),\qquad d=\Theta(\ell),
 \label{eqn:almost-scales}
\end{equation}
which yields Equation~\eqref{eqn:almost-linear}.
The matrix expansion assumption is written out in Listing~\ref{lst:pk20-premise}.

\subsection{Asymptotically good lifted product codes}
\label{sec:good-lifted}

Good lifted product codes~\cite{PK22} achieve linear dimension and linear distance simultaneously over every finite field.
Their local codes satisfy a product expansion condition that controls decompositions into contributions from the two factors.
This additional control allows local reductions to be used throughout the global complex.
The local matrices and graph degree are fixed before the graph size grows, so the resulting rate, relative distance and locality bounds are uniform along the family.

\begin{theorem}[Asymptotically good lifted product codes~\cite{PK22}]
\label{thm:good-lifted}
Let $F$ be a finite field and $0<R<1$.
Using the LPS spectral estimate, the construction gives a family over $F$ with unbounded length and
\begin{equation}
  k\ge Rn,\qquad d\ge\delta n
  \label{eqn:pk-good}
\end{equation}
for some fixed $\delta>0$ and all sufficiently large members.
Check weights and incidences are uniformly bounded.
\end{theorem}

Listing~\ref{lst:pk-good} realizes Theorem~\ref{thm:good-lifted} using Definition~\ref{def:good-family}.
The outputs \lean{h} and \lean{h'} are the local matrices, and \lean{a} fixes their compatible degree.
Here $p$ is the prime parameter of the LPS construction.
For degree $p+1$, the spectral bound in the input is $2\sqrt p$, as in Equation~\eqref{eqn:lps-spectrum}.

\begin{leanlisting}{Theorem~\ref{thm:good-lifted} in Lean}{lst:pk-good}
-- QEC.Coding.ExpanderCodes.LPSGoodFamilyAssembly:415-432
theorem exists_pk22LPSFamily_of_ramanujanComplexOperatorBound
    (F : Type) [Field F] [Fintype F]
    (targetRate : ℝ) (hRatePositive : 0 < targetRate)
    (hRate : targetRate < 1)
    (hRamanujan : ∀ (p : ℕ) (hp : p.Prime) (hp4 : p % 4 = 1) (n : ℕ),
      SmallSetExpansion.HasMeanZeroComplexOperatorBound
        (rightCayleyGraph (lpsQuaternionFamilyLabel hp hp4 n))
        (2 * Real.sqrt (p : ℝ))) :
    ∃ (t : ℕ)
      (a : CompatibleDegreePrime (pk22LocalWidth (Nat.card F) t))
      (h h' : Matrix (PK22Check (Nat.card F) t a)
        (PK22Direction (Nat.card F) t a) F),
      let hCount := card_lpsQuaternionShell_eq_prime_add_one
        a.prime a.oneModFour
      QEC.QLDPC.IsAsymptoticallyGoodQLDPCFamily
          (lpsSemanticFamily a.prime a.oneModFour hCount h h') ∧
        QEC.QLDPC.HasEventualRateAtLeast
          (lpsSemanticFamily a.prime a.oneModFour hCount h h') targetRate
\end{leanlisting}

For the distance argument, choose a representative of a nontrivial homology class with minimum weight under addition of boundaries.
No local boundary addition can reduce its weight, so it is also locally minimal.
The combination of local product expansion and graph expansion forces any nonzero locally minimal cycle to occupy a fixed fraction of the physical coordinates~\cite{PK22}.
This gives the distance bound, while the choice of local matrix dimensions controls the rate.

\subsection{Quantum Tanner codes}
\label{sec:tanner}

For the quantum Tanner construction of Section~\ref{sec:tanner-construction}, the distance proof combines expansion of the two Cayley graphs with robustness of the local tensor codes~\cite{LZ22}.
Classical distance alone does not describe all the cancellations that can occur when row and column contributions overlap.
Robustness controls this additional issue by ensuring that a small word in the tensor sum can be confined to a suitably small collection of rows and columns.
This control is used to reduce the weight of a logical representative by local stabilizers.

\begin{definition}[Robustness of local tensor codes]
\label{def:robustness}
For the local codes $C_A\subseteq\mathbb F_2^A$ and $C_B\subseteq\mathbb F_2^B$ of Definition~\ref{def:local-tensor}, write
\begin{equation}
 C_A\boxplus C_B=(C_A\otimes\mathbb F_2^B)+(\mathbb F_2^A\otimes C_B).
 \label{eqn:tensor-sum}
\end{equation}
The pair is robust up to weight $w$ if every $c\in C_A\boxplus C_B$ with $\operatorname{wt}(c)\le w$ is supported on a union of rows indexed by $R\subseteq A$ and columns indexed by $S\subseteq B$ satisfying
\begin{equation}
 |R|\le\frac{\operatorname{wt}(c)}{d_B},\qquad
 |S|\le\frac{\operatorname{wt}(c)}{d_A},
 \label{eqn:robustness}
\end{equation}
where $d_A,d_B>0$ are the local code distances.
\end{definition}

Listing~\ref{lst:robustness} is the support formulation used in the distance proof.
The construction selects local codes for which the required robustness bounds also survive the specified punctures~\cite{LZ22}.
Graph expansion controls how small exceptional neighborhoods meet, while local robustness permits a reduction of a low weight representative by stabilizers.
These two ingredients exclude nontrivial logical operators below the distance bound~\cite{LZ22}.

\begin{leanlisting}{Definition~\ref{def:robustness} in Lean}{lst:robustness}
-- QEC.Coding.QuantumTanner.Core
def IsRobust {A B : Type v} [Fintype A] [Fintype B]
    (CA : LinearCode (ZMod 2) A) (CB : LinearCode (ZMod 2) B) (w : Nat) : Prop :=
  ∀ dA dB, MinimumDistance CA dA → MinimumDistance CB dB →
    ∀ x, x ∈ (dualTensorCode CA CB).carrier → HammingWeight x ≤ w →
      ∃ rows : Finset A, ∃ columns : Finset B,
        SupportCovered x rows columns ∧
          rows.card ≤ HammingWeight x / dB ∧
          columns.card ≤ HammingWeight x / dA
\end{leanlisting}

Let $h_2$ denote binary entropy with logarithms to base two.

\begin{theorem}[Asymptotically good quantum Tanner codes~\cite{LZ22}]
\label{thm:tanner}
Fix $0<\rho,\epsilon,\delta<1/2$ with $h_2(\delta)<\rho$.
Here $\rho$ controls the complementary local code rates.
The parameter $\delta$ bounds the relative distances of the local codes and their duals from below, and $\epsilon$ provides slack in the distance exponent.
Choose a sufficiently large admissible common Cayley graph degree $\Delta=|A|=|B|$ and hold it fixed.
Assuming the spectral bounds for both Cayley graph families, a fixed choice of local codes gives a binary qLDPC family with
\begin{equation}
  k\ge(1-2\rho)^2n,\qquad
  d\ge\frac{\delta}{4\Delta^{3/2+\epsilon}}n.
  \label{eqn:tanner-parameters}
\end{equation}
The chosen check weights and qubit incidences are at most $\Delta^2$.
A uniformly sampled pair of local matrices has positive probability of giving such a family.
\end{theorem}

Listing~\ref{lst:tanner} formalizes Theorem~\ref{thm:tanner} for the chosen arithmetic graph sequence.
A single successful sample of the local codes works for the entire sequence; the local codes are chosen once and then held fixed as the graph grows.

\begin{leanlisting}{Theorem~\ref{thm:tanner} in Lean}{lst:tanner}
-- QEC.Results.QLDPC.ConditionalQuantumTannerFamily:41-68
theorem eventually_positiveProbability_good_arithmetic_code_family_of_complex_spectral_bounds
    (delta eta rho : ℝ) (hDelta : 0 < delta) (hDeltaHalf : delta < 1 / 2)
    (hEta : 0 < eta) (hEtaHalf : eta < 1 / 2)
    (hRho : 0 < rho) (hRhoHalf : rho < 1 / 2)
    (hEntropy : Real.binEntropy delta < rho * Real.log 2) :
    ∃ t : ℕ, ∀ (p : ℕ) [Fact p.Prime] (hOdd : 3 ≤ p)
      (epsilon : ZMod p) (hEpsilon : epsilon ≠ 0),
      (∀ r : ZMod p, r ^ 2 ≠ epsilon) → t ≤ p + 1 →
      (∀ n : ℕ,
        HasMeanZeroComplexOperatorBound
          (squareComplex p hOdd epsilon hEpsilon n).leftCayleyGraph
          (2 * Real.sqrt ((p + 1 : ℕ) : ℝ)) ∧
        HasMeanZeroComplexOperatorBound
          (squareComplex p hOdd epsilon hEpsilon n).rightCayleyGraph
          (2 * Real.sqrt ((p + 1 : ℕ) : ℝ))) →
      HasPositiveUniformProbability (fun sample =>
        HasGoodSampleFamily (left p hOdd epsilon hEpsilon) (right p hOdd epsilon hEpsilon)
          (indexedLeftInverse epsilon hEpsilon) (Equiv.refl _)
          (left_injective p hOdd epsilon hEpsilon) (right_injective p hOdd epsilon hEpsilon)
          (left_inverse p hOdd epsilon hEpsilon) (right_inverse p hOdd epsilon hEpsilon)
          (total_nonconjugacy p hOdd epsilon hEpsilon) delta eta rho (p + 1) sample ∧
        letI : ∀ n, DecidableEq (CayleySquare
            (left p hOdd epsilon hEpsilon n) (right p hOdd epsilon hEpsilon n)) :=
          fun _ => Subtype.instDecidableEq
        let family := sampleFamily p hOdd epsilon hEpsilon rho sample
        ∃ hCofinal : ∀ N : ℕ, ∃ n, N ≤ (family n).blockLength,
          _root_.QEC.QLDPC.IsAsymptoticallyGoodQLDPCFamily
            (Family.ofData family hCofinal))
\end{leanlisting}

The Lean parameter \lean{eta} is the exponent $\epsilon$ in Equation~\eqref{eqn:tanner-parameters}; \lean{epsilon} in the code denotes a nonsquare field element in the arithmetic construction.
The condition \lean{Real.binEntropy delta < rho * Real.log 2} is the natural logarithm form of $h_2(\delta)<\rho$.
The restriction $\delta<1/2$ selects the branch of the entropy function used in the source argument~\cite{LZ22}.

Theorem~\ref{thm:good-lifted} and Theorem~\ref{thm:tanner} both establish asymptotic goodness in the sense of Definition~\ref{def:good-family}.
\section{Further Applications}
\label{sec:applications}

Beyond asymptotic code parameters, QEC theory studies the physical constraints and resource costs of protecting quantum information.
This section presents formal results on geometric locality, magic state distillation, and passive quantum memories.
These developments combine code properties with spatial constraints, protocol descriptions, and models of thermal dynamics.

\subsection{Locality and connectivity bounds}
\label{sec:locality}

The qLDPC condition bounds the number of qubits touched by each check and the number of checks touching each qubit.
It leaves open how far apart those qubits are in a spatial realization.
This distinction matters for the codes of Section~\ref{sec:qldpc}, whose favorable parameters need not be compatible with interactions of bounded range in two dimensions.
Locality bounds quantify this restriction and connect abstract code parameters to the geometry of a possible implementation~\cite{BPT10,DL24}.

The Bravyi--Poulin--Terhal bound~\cite{BPT10} limits the parameters of codes defined by commuting local projectors in two dimensions.
Let $C$ be their common eigenspace on a square lattice of $n$ sites, with local Hilbert space dimension at most $q$.
Write $k=\log_2\dim C$.
A distance $d$ means that every measurement supported on fewer than $d$ sites reveals no information about the encoded state~\cite{BPT10}.

\begin{theorem}[BPT bound~\cite{BPT10}]
\label{thm:bpt}
Suppose each projector is supported in a square of side at most $w$ and $d>0$.
There is a constant $C(w,q)$, independent of the lattice size, such that
\begin{equation}
  kd^2\le C(w,q)n.
  \label{eqn:bpt-finite}
\end{equation}
\end{theorem}

Listing~\ref{lst:bpt} is the finite lattice form of Theorem~\ref{thm:bpt}.
The Lean statement takes $C(w,q)=16384\,w^4\log_2q$.
The variables \lean{M} and \lean{parameter} specify the commuting projectors and encoded dimension.
The predicate \lean{BPTMeasurementDistance} expresses the measurement condition above.

\begin{leanlisting}{Theorem~\ref{thm:bpt} in Lean}{lst:bpt}
-- QEC.Results.LocalityBounds.GeometricTradeoff:33-42
theorem bpt_geometric_locality_tradeoff
    (M : BPTCommutingProjectorModel L kappa)
    (parameter : BPTCommutingProjectorModel.EncodedQubitParameter M)
    (distance : Nat)
    (hDistance : Semantics.QuditSubsystem.BPTMeasurementDistance M.codeSpace distance)
    (hdistance : 0 < distance)
    (hregular : L.lattice.rows = L.lattice.columns) :
    (parameter.encodedQubits : Real) * (distance : Real) ^ 2 <=
      (16384 : Real) * (M.interactionRange : Real) ^ 4 *
        Real.logb 2 (L.uniformDimensionBound : Real) * (L.lattice.siteCount : Real)
\end{leanlisting}

The proof first translates the absence of information in a region into correctability of its erasure.
A disentangling lemma then uses the correctability of the region and its external boundary to separate a fixed interior state from the remaining encoded information~\cite{BPT10}.
Locality allows correctable regions to be enlarged and combined into a geometric partition.
An entropy bound on the remaining boundary region limits the encoded dimension; choosing the partition scale in terms of $d$ gives Equation~\eqref{eqn:bpt-finite}.
The formal proof constructs both the partition and the required recovery maps.

Connectivity bounds extend this viewpoint to the interaction graph of a chosen stabilizer presentation.
Baspin--Krishna~\cite{BK22} relate code parameters to the sizes of graph separators, and BGKL~\cite{BGKL24} obtain improved rate--distance bounds through a refined partition argument.
The library includes binary stabilizer versions of both results, linking recursive graph partitions to correctable regions and encoded dimension bounds.

A related result concerns the interactions required when stabilizer codes are embedded in the plane.
For a planar embedding with minimum separation one, the Dai--Li bound~\cite{DL24} implies that, once $kd^2\ge c_1n$, there are at least $c_0\max(k,d)$ distinct interactions of length at least
\begin{equation}
  c_0\max\!\left\{\frac{d}{\sqrt n},
                   \left(\frac{kd^2}{n}\right)^{1/4}\right\},
  \label{eqn:long-interaction-scale}
\end{equation}
where $c_0,c_1>0$ are absolute constants.
The interactions are determined by the chosen stabilizer generators.
This theorem is formalized for that generating presentation, whereas Theorem~\ref{thm:bpt} applies to commuting projectors on a geometric lattice.

\subsection{Magic state distillation and transversal gates}
\label{sec:distillation}

Magic state distillation uses error correction to convert noisy resource states into states of higher fidelity.
Its performance is measured by the output error and the cost of producing successful outputs.
The overhead describes how many noisy inputs are consumed per output as the target error decreases.

The Bravyi--Kitaev protocols~\cite{BK05} establish input fidelity thresholds above which noisy ancillas can be distilled to support universal quantum computation with ideal Clifford operations.
Triorthogonal codes~\cite{BH12} subsequently provided protocols with lower resource overhead.
Both the threshold results and these overhead guarantees are included in the library.

Further constructions reduce the overhead below logarithmic growth in the inverse target error~\cite{HH18}, or reduce the space needed while controlling the asymptotic input count~\cite{HHPW17}.
Protocols using qudits of prime dimension~\cite{KT19} offer overhead exponents approaching zero as that dimension increases.
The resource state changes with the dimension in this limit.
The library formalizes the sublogarithmic and prime-qudit results under their respective noise and resource conventions.
These developments motivate the stronger goal of a bounded input cost per output at arbitrarily small target error.

\paragraph{Constant overhead under local stochastic noise.}
The constant overhead construction~\cite{WHY24} achieves this goal using a code family with positive rate and growing error-correcting capability.
It also allows correlated faults through a local stochastic noise model.
If $S$ is the random set of faulty input locations, the local stochastic condition with rate $p$ is
\begin{equation}
  \Pr(T\subseteq S)\le p^{|T|}
  \qquad\text{for every set of locations }T.
  \label{eqn:local-stochastic}
\end{equation}
Quantum errors conditioned on $S$ may be correlated within their support~\cite{WHY24}.
This condition bounds the probability of a specified collection of simultaneous faults, even when the complete fault distribution does not factor into independent probabilities.
The Lean preparation type \lean{Why24LocalStochasticPreparation} includes this condition.

The protocol uses a code over $\mathbb F_{1024}$ and a fixed conversion between binary controlled-controlled-$Z$ (CCZ) resources and qudit resources~\cite{WHY24}.
Let $r\in\mathbb N$ index the selected code and protocol sequence.
Let $c_{\rm conv}$ be the number of binary CCZ resources used per qudit resource, and let $n_r,k_r,t_r$ denote code length, output count and decoding radius.
Write $\rho_{\rm out}$ for the output state and $\rho_{\rm ideal}$ for its ideal target.
We measure output error by the normalized trace distance $\epsilon_{\rm out}=\tfrac12\|\rho_{\rm out}-\rho_{\rm ideal}\|_1$, with $\|M\|_1=\operatorname{Tr}\sqrt{M^\dagger M}$.

\begin{theorem}[One round of distillation]
\label{thm:distillation-round}
For an input preparation satisfying Equation~\eqref{eqn:local-stochastic}, the protocol accepts with probability one, consumes $N_{\rm in}=c_{\rm conv}n_r$ resources, and produces $N_{\rm out}=k_r$ outputs with error
\begin{equation}
  \epsilon_{\rm out}(r)
  \le {n_r\choose t_r+1}
       p_{\rm eff}^{\,t_r+1},
  \qquad p_{\rm eff}=c_{\rm conv}p.
  \label{eqn:distillation-error}
\end{equation}
\end{theorem}

Decoding returns the intended logical resource whenever at most $t_r$ qudit locations are faulty.
The output error is therefore bounded by the probability of more faults.
Every larger fault support contains a set of $t_r+1$ locations; applying Equation~\eqref{eqn:local-stochastic} and summing over these sets gives Equation~\eqref{eqn:distillation-error}.
The fixed conversion in the formal construction has $c_{\rm conv}=70$~\cite{WHY24}.
Listing~\ref{lst:distillation-round} uses this value and proves the resource count, acceptance probability and error bound for the same physical protocol.

\begin{leanlisting}{Theorem~\ref{thm:distillation-round} in Lean}{lst:distillation-round}
-- QEC.Distillation.ConstantOverheadProtocol.OneRound:5794-5811
theorem why24_one_round_physical_protocol (r : Nat)
    (preparation : Why24LocalStochasticPreparation
      (Fin (why24OneRoundInputCount r))
      (why24IdealCCZRegisterDensity (why24OneRoundInputCount r))) :
    let protocol :=
      why24OneRoundMagicStateDistillationProtocol r preparation
    protocol.run = why24OneRoundRun r preparation ∧
      protocol.inputErrorModel = InputErrorModel.localStochastic
        (why24LocalStochasticInputErrorModel preparation) ∧
      protocol.behavior = .activeCorrection (QuantumChannel.id _) ∧
      protocol.run.resourceCount = 70 * f1024Why24OddLength r ∧
      why24OneRoundOutputCount r = f1024Why24OddDimension r ∧
      protocol.run.successProbability = 1 ∧
      protocol.successfulOutputError ≤
        (Nat.choose (f1024Why24OddLength r)
          (f1024Why24OddDecodingRadius r + 1) : Real) *
          (70 * preparation.rate) ^
            (f1024Why24OddDecodingRadius r + 1)
\end{leanlisting}

The selected algebraic geometric code family has positive rate and a decoding radius proportional to its length~\cite{WHY24}.
In the error bound of Theorem~\ref{thm:distillation-round}, the binomial factor grows at most exponentially with block length.
For sufficiently small fixed $p$, the factor $p_{\rm eff}^{\,t_r+1}$ decays faster because $t_r$ grows linearly with $n_r$.
The output error consequently tends to zero.
Positive rate bounds $n_r/k_r$, so including the fixed conversion cost still gives a bounded number of inputs per output.

\begin{theorem}[Constant overhead distillation~\cite{WHY24}]
\label{thm:constant-overhead}
There are constants $p_0,\Omega>0$ such that, for any physical rate $0\le p<p_0$ and target error $\epsilon>0$, all sufficiently large members of the protocol family satisfy
\begin{equation}
  \epsilon_{\rm out}\le\epsilon,\qquad
  \frac{N_{\rm in}}{N_{\rm out}}\le\Omega.
  \label{eqn:constant-overhead}
\end{equation}
The estimates hold uniformly over input preparations with local stochastic rate at most $p$.
\end{theorem}

Listing~\ref{lst:distillation-full} in Appendix~\ref{appendix:applications-statements} gives the complete Lean statement of Theorem~\ref{thm:constant-overhead}, including finite resource conversions and catalyst accounting.
The catalyst has a positive initialization cost, whose contribution per fresh output tends to zero as the output count grows.

The overhead here counts resource states under the specified noise model.
For protocols with postselection, this cost must account for unsuccessful rounds; the protocol in Theorem~\ref{thm:distillation-round} accepts with probability one.

The role of good codes extends to the logical gates available on encoded states.
Related constructions support transversal non-Clifford gates while retaining linear dimension and distance~\cite{GG24,Ngu24}.
For the binary construction of Ref.~\cite{Ngu24}, the library proves these parameter guarantees and verifies a physical CCZ operation implementing parallel logical CCZ gates across three encoded blocks.
The classical codes underlying these constructions must combine suitable rate and distance with the algebraic properties needed for transversal gates.
The supporting algebraic geometric development includes function-field towers~\cite{GS95,GS05} and code duality~\cite{Sti09}.
In this setting, self-orthogonal and self-dual codes can also attain the Tsfasman--Vladut--Zink bound~\cite{Sti06}.

\subsection{Energy barriers and quantum memories}
\label{sec:memory}

Passive quantum storage asks whether a Hamiltonian and its coupling to a thermal environment can preserve encoded information for a long time without repeated active correction.
We first examine the energetic cost of building a logical error~\cite{BT09}, then consider how thermal probabilities and dynamics control the storage error at finite times~\cite{CLBT10}.

\paragraph{Logical paths and energy barriers.}
Let $\{S_a\}_{a\in\mathcal I}$ be the chosen commuting Hermitian stabilizer generators, with unit coupling coefficients in $H=-\sum_{a\in\mathcal I}S_a$.
Relative to their common $+1$ eigenspace, a Pauli error $E$ has excess energy
\begin{equation}
  \varepsilon_H(E)=2\,|\{a\in\mathcal I:ES_a=-S_aE\}|.
  \label{eqn:pauli-energy}
\end{equation}
A Pauli path $\gamma=(E_0,\ldots,E_m)$ with $m$ steps starts at the identity and changes by an operator on at most one qubit at each step.
Its height is $\max_j\varepsilon_H(E_j)$.
Minimizing this height over paths ending in a specified logical sector defines the sector energy barrier~\cite{BT09,DKLP02}.

In two dimensions, the Bravyi--Terhal argument~\cite{BT09} bounds the logical Pauli barrier by a constant when interaction range and generator participation are bounded.
The formal theorem covers the chosen local stabilizer Hamiltonians on open and periodic square lattices.
The proof uses cleaning to find a logical operator supported in a narrow strip.
Its single-qubit steps can be ordered so that only a bounded number of checks is violated at once~\cite{BT09}.
The logical operator may have large weight, but the maximum energy encountered along this path stays bounded.
This illustrates why distance and energy barrier describe different aspects of protection.

The toric code in four dimensions~\cite{DKLP02} gives a contrasting example.
Qubits occupy plaquettes of a periodic lattice of side length $L$, and nontrivial membranes represent logical operators.
As a membrane is grown, excitations occur along its boundary~\cite{DKLP02}.
A completed nontrivial membrane returns to zero syndrome, but an intermediate configuration must carry a boundary whose size grows with the lattice.
Write $b_L(s)$ for the barrier in a nonzero coordinate membrane sector $s$, in either the primal or dual description.

\begin{theorem}[Linear barriers in coordinate membrane sectors~\cite{DKLP02}]
\label{thm:toric-barrier}
There are constants $c,C>0$, independent of $L$ and $s$, such that for every integer $L>1$ and every such sector,
\begin{equation}
  cL\le b_L(s)\le CL.
  \label{eqn:toric-barrier}
\end{equation}
\end{theorem}

Listing~\ref{lst:toric-barrier} proves Theorem~\ref{thm:toric-barrier} by bounding the height of every path to the sector from below and constructing a membrane sweep for the upper bound.
The Lean bounds $\lfloor L/2\rfloor$ and $384L$ permit $c=1/3$ and $C=384$ for integer $L>1$.

\begin{leanlisting}{Theorem~\ref{thm:toric-barrier} in Lean}{lst:toric-barrier}
-- QEC.Results.QuantumMemory.FourDimensionalEnergyBarrier:37-51
theorem fourDimensionalToric_linearEnergyBarrier_theta :
    (∀ (s : MembraneSector) (γ : PauliErrorPath (ι := Plaquette L)),
      HasPrimalSectorEndpoint γ s →
        L / 2 ≤ (toricHamiltonian (L := L)).pathHeight γ) ∧
    (∀ (s : MembraneSector) (γ : PauliErrorPath (ι := Plaquette L)),
      HasDualSectorEndpoint γ s →
        L / 2 ≤ (toricHamiltonian (L := L)).pathHeight γ) ∧
    (∀ s : MembraneSector,
      HasPrimalSectorEndpoint (primalSectorSweepPath (L := L) s) s ∧
        (toricHamiltonian (L := L)).pathHeight
          (primalSectorSweepPath (L := L) s) ≤ 384 * L) ∧
    (∀ s : MembraneSector,
      HasDualSectorEndpoint (dualSectorSweepPath (L := L) s) s ∧
        (toricHamiltonian (L := L)).pathHeight
          (dualSectorSweepPath (L := L) s) ≤ 384 * L)
\end{leanlisting}

\paragraph{Thermal protection and decoding.}
To obtain a storage time, one must also control the dynamics at finite temperature.
At nonzero temperature, energetic suppression competes with the number of configurations that can realize an error.
A storage estimate therefore needs a bound on the thermal weight of configurations where logical protection can fail~\cite{CLBT10}.
The protected observable method~\cite{CLBT10} expresses this requirement through a protected subspace and relates the storage error to the Gibbs weight outside it.
Let $\widetilde X,\widetilde Z$ be Hermitian logical observables satisfying $\widetilde X^2=\widetilde Z^2=I$ and $\widetilde X\widetilde Z=-\widetilde Z\widetilde X$.
Let $P$ be an orthogonal projection and $J_a$ the Lindblad jump operators, with $a$ running over their finite index set.
At inverse temperature $\beta$, write $\rho_\beta=e^{-\beta H}/\operatorname{Tr}(e^{-\beta H})$.
The dissipator is $\mathcal L_{\rm diss}(\rho)=\sum_a(J_a\rho J_a^\dagger-\tfrac12\{J_a^\dagger J_a,\rho\})$, where $\{U,V\}=UV+VU$.
The full generator is $\mathcal L(\rho)=-i[H,\rho]+\mathcal L_{\rm diss}(\rho)$, with $[U,V]=UV-VU$.
Assume $H$ commutes with both observables, $P$ commutes with $H$ and the observables, and
\begin{equation}
  [J_a,\widetilde X]P=[J_a,\widetilde Z]P=0
  \qquad\text{for every }a.
  \label{eqn:jump-protection}
\end{equation}

\begin{theorem}[Protected observables~\cite{CLBT10}]
\label{thm:protected-decay}
Suppose the Gibbs state $\rho_\beta$ is stationary for the specified dissipator.
There are CPTP maps $\operatorname{Enc}$ and $\operatorname{Dec}$, with
$\operatorname{Dec}\circ\operatorname{Enc}=\operatorname{id}$ on states of a single qubit, such that
\begin{equation}
  \bigl\|\Phi_t(\operatorname{Enc}(\eta))-
               \operatorname{Enc}(\eta)\bigr\|_1
  \le C_{\rm prot}\,t\,
       \|\mathcal L_{\rm diss}\|_{1\to1,h}\,
       \operatorname{Tr}[(I-P)\rho_\beta],
  \qquad t\ge0,
  \label{eqn:protected-decay}
\end{equation}
for every density operator $\eta$ of a single qubit, with a universal constant $C_{\rm prot}>0$.
Here $\Phi_t=e^{t\mathcal L}$ is the physical Lindblad evolution at time $t$, and the dissipator norm is induced by the trace norm on Hermitian operators.
\end{theorem}

Listing~\ref{lst:protected-decay} establishes Theorem~\ref{thm:protected-decay} with $C_{\rm prot}=8$.
The factor $\operatorname{Tr}[(I-P)\rho_\beta]$ is the equilibrium probability of being outside the protected subspace.
In the proof, jump protection and Gibbs stationarity bound the rate of change of the encoded state by this probability times the dissipator norm~\cite{CLBT10}.
Integration gives the factor of $t$, and contraction under the decoding channel gives the same error bound for the stored qubit.

\begin{leanlisting}{Theorem~\ref{thm:protected-decay} in Lean}{lst:protected-decay}
-- QEC.Semantics.Dynamics.ProtectedObservableDecay:623-650
theorem clbtProtectedObservableDecay
    {K ι : Type}
    [NormedAddCommGroup K] [InnerProductSpace ℂ K] [FiniteQuantumSystem K]
    [Nontrivial K] [Fintype ι]
    (logical : ThermalEncoding.PauliLikePair K)
    (hamiltonian : Operator K) (hhamiltonian : hamiltonian.IsSymmetric)
    (beta : ℝ) (J : ι → Operator K) (P : Operator K)
    (hP : IsStarProjection P)
    (hPx : Commute P logical.x) (hPz : Commute P logical.z)
    (hHx : Commute hamiltonian logical.x)
    (hHz : Commute hamiltonian logical.z)
    (hPH : Commute P hamiltonian)
    (hprotectedX : JumpProtected J logical.x P)
    (hprotectedZ : JumpProtected J logical.z P)
    (hstationary : lindbladDissipator J
      (logical.physicalGibbsState hamiltonian hHz hhamiltonian beta).toOperator = 0) :
    ∃ Enc : QuantumChannel Qubit.QubitSpace K,
      ∃ Dec : QuantumChannel K Qubit.QubitSpace,
        (∀ eta : DensityOperator Qubit.QubitSpace,
          Dec.mapDensity (Enc.mapDensity eta) = eta) ∧
        ∀ (eta : DensityOperator Qubit.QubitSpace) (t : ℝ) (ht : 0 ≤ t),
          traceNorm
              (((finiteLindbladEvolutionChannel hamiltonian J hhamiltonian t ht)
                  |>.mapDensity (Enc.mapDensity eta)).toOperator -
                (Enc.mapDensity eta).toOperator) ≤
            8 * t * lindbladDissipatorNorm J *
              clbtGibbsTailMass P
                (logical.physicalGibbsState hamiltonian hHz hhamiltonian beta)
\end{leanlisting}

For stabilizer Hamiltonians, \lean{stableSyndromeLifetimeBridge} uses a decoder to construct protected logical observables and bounds the storage error by the Gibbs weight of bad syndromes.
The library also proves Gibbs stationarity for Davies dynamics with transition rates satisfying detailed balance.
For a particular memory, one must verify that its bath jumps preserve the protected observables and bound the Gibbs weight of bad syndromes.

\paragraph{Toward a three-dimensional passive memory.}
The two-dimensional restrictions~\cite{BT09} and the four-dimensional protection mechanism~\cite{DKLP02} motivate the search for a passive memory in three spatial dimensions.
For many years, it remained open whether a local three-dimensional system could preserve quantum information at a fixed nonzero temperature for a time that grows without bound with its size~\cite{BH13,BDL26}.
Such a memory must protect both logical $X$ and $Z$ information and control the thermal fluctuations that can damage it.
Requiring local interactions in three spatial dimensions brings this question closer to the geometry of physical devices.

The construction in Ref.~\cite{BDL26} uses three-dimensional local interactions with bounded range and density.
At sufficiently low temperature, it gives a lower bound on storage time that grows as a stretched exponential of the number of qubits under the specified thermal dynamics~\cite{BDL26}.
Its recursive code construction is combined with a multiscale estimate suppressing the Gibbs weight of unstable syndromes.
The protected-observable method then converts this probability estimate into a storage guarantee~\cite{BDL26}.
The library currently establishes the protected-observable theorem, its specialization to stabilizer syndromes, and supporting results on thermal dynamics.
The concrete three-dimensional construction and its lifetime bound remain subsequent goals.
\section{Conclusion}
\label{sec:conclusion}

We have presented QECert, a Lean library for formalizing the asymptotic properties and fundamental limits of quantum error correction.
Its common mathematical framework connects quantum descriptions of error correction with the algebraic and homological methods used to study quantum codes.
The principal case study covers major developments in qLDPC constructions, culminating in asymptotically good lifted product and quantum Tanner codes.
Further results address locality restrictions, constant overhead magic state distillation, and quantum memories.
Together, these developments bring central theoretical guarantees concerning error protection, resource efficiency, and physical constraints into a reusable body of verified mathematics.

QECert makes the assumptions behind these guarantees explicit and provides common definitions through which results from different constructions can be related.
It also helps distinguish properties specific to a construction from general statements that apply across broader classes of quantum codes.
This supports the cumulative development of QEC theory, allowing new proofs to build on established results.

Future work can extend this foundation to additional code families, decoding guarantees, and fault-tolerant protocols, connecting more of coding theory to the procedures through which protection is realized.
As AI agents take part in mathematical discovery, rigorous verification will be essential to establishing new theoretical results in QEC.
We hope that QECert will contribute to a shared digital foundation on which researchers and AI systems can develop the theory of quantum error correction together.

\medskip
\noindent\textit{Acknowledgment.}---This work was partially supported by the National Natural Science Foundation of China (Grant. No. 92576114, 12447107), the Guangdong Provincial Quantum Science Strategic Initiative (Grant No. GDZX2403008, GDZX2503001).

\clearpage
\appendix
\section{Supplementary Statements for Code Constructions and Distillation}
\label{appendix:statements}

This appendix supplements the constructions in Section~\ref{sec:qldpc} with finite parameter certificates and a matrix expansion assumption, and explains the resource accounting in Theorem~\ref{thm:constant-overhead}.
Lean listings omit proofs, imports and surrounding section variables.

\subsection{Supplementary statements for qLDPC constructions}
\label{appendix:qldpc-statements}

\subsubsection{Finite parameter bounds for balanced product codes}
\label{appendix:balanced-statements}

The finite certificates returned by Theorem~\ref{thm:balanced} imply its two asymptotic parameter regimes.
As in the main text, $n$ is the physical length and $k$ is the number of logical qubits.
The quantities $\underline d_X,\underline d_Z$ are certified lower bounds on the dressed distances, so $\underline d_X\le d_X$ and $\underline d_Z\le d_Z$.
All coefficients are positive constants independent of the admissible prime $q$ indexing the construction.

\paragraph{Asymmetric distances.}
The first construction satisfies
\begin{align}
 n^2&\le a_-k^3, & k^3&\le a_+n^2,\\
 n&\le a_X\underline d_X^3, & n&\le a_Z\underline d_Z,\qquad \underline d_Z\le n.
 \label{eqn:unbalanced-finite}
\end{align}
Here $a_-,a_+$ control the number of logical qubits and $a_X,a_Z$ control the distance lower bounds.
The first line gives $k=\Theta(n^{2/3})$, while the second gives $d_X=\Omega(n^{1/3})$ and $d_Z=\Omega(n)$.
Since a nontrivial logical operator has weight at most $n$, the latter bound yields $d_Z=\Theta(n)$.
These are the parameters in Equation~\eqref{eqn:balanced-unbalanced}.

Listing~\ref{lst:be-unbalanced-certificate} also specifies the exact length $n=401q(q^2-1)$ and a uniform locality bound of $404$.
The Lean identifiers \lean{N} and \lean{K} correspond to $n$ and $k$, respectively.
The identifiers \lean{dX} and \lean{dZ} denote the lower bounds $\underline d_X,\underline d_Z$.
The initial construction \lean{D} selects the horizontal logical subsystem of the cyclic product.

\begin{leanlisting}{Finite certificate for the asymmetric distances in Theorem~\ref{thm:balanced}}{lst:be-unbalanced-certificate}
-- QEC.Coding.BalancedProduct.UnbalancedFamily
def UnbalancedFamilyCertificate
    (q : Nat) [Fact q.Prime]
    (hq40 : 40 < q)
    (hq : legendreSym q 401 = -1) : Prop :=
  let hOdd : Odd q :=
    (Fact.out : q.Prime).odd_of_ne_two (by omega)
  let D := HorizontalSubsystem.horizontalSubsystem
    (lpsTannerGroupAlgebraMatrix q hq40 hq)
    (Cyclic.cyclicGenerator q)
    (by simpa [Cyclic.CyclicGroup, ZMod.card] using hOdd)
  let N := Fintype.card
    (LiftedProduct.QubitIndex
      (CheckOrbit401 q) (EdgeOrbit401 q hq) Unit Unit
      (Cyclic.CyclicGroup q))
  let K := D.logicalQubits
  let dX := lpsDressedXLowerBound q
  let dZ := lpsDressedZLowerBound q hq
  (D.complex.binaryCSSCode).stabilizer.IsSourceValidBinary ∧
    D.DressedXDistanceAtLeast dX ∧
    D.DressedZDistanceAtLeast dZ ∧
    BoundedLocality
      (LiftedProduct.boundaryIncidence
        (Cyclic.liftedProductData
          (lpsTannerGroupAlgebraMatrix q hq40 hq)
          (Cyclic.cyclicGenerator q))).localityProfile 404 ∧
    N = 401 * q * (q ^ 2 - 1) ∧
    q ^ 2 - 1 ≤ K ∧
    K ≤ 201 * (q ^ 2 - 1) ∧
    N ^ 2 ≤ dimensionLowerCoefficient * K ^ 3 ∧
    K ^ 3 ≤ dimensionUpperCoefficient * N ^ 2 ∧
    N ≤ dressedXCoefficient * dX ^ 3 ∧
    N ≤ dressedZCoefficient * dZ ∧
    dZ ≤ N
\end{leanlisting}

\paragraph{Balanced distances.}
The second construction satisfies
\begin{equation}
 n^4\le c_-k^5,\qquad k^5\le c_+n^4,\qquad
 n^3\le c_X\underline d_X^5,\qquad n^3\le c_Z\underline d_Z^5.
 \label{eqn:balanced-finite}
\end{equation}
The constants $c_-,c_+$ control the number of logical qubits and $c_X,c_Z$ control distance.
Taking the corresponding roots gives $k=\Theta(n^{4/5})$ and
$d=\min(d_X,d_Z)=\Omega(n^{3/5})$, as in Equation~\eqref{eqn:balanced-family}.
Listing~\ref{lst:be-certificate} records these inequalities with the exact length and a uniform locality bound of $806$.
The symbols \lean{N}, \lean{K}, \lean{dX} and \lean{dZ} have the same roles as in the preceding certificate.

\begin{leanlisting}{Finite certificate for the balanced distances in Theorem~\ref{thm:balanced}}{lst:be-certificate}
-- QEC.Coding.BalancedProduct.BalancedFamilyLocality:148-168
def BalancedFamilyCertificate
    (q : Nat) [Fact q.Prime]
    (hq40 : 40 < q) (hq : legendreSym q 401 = -1) : Prop :=
  let D := balancedSubsystem q hq40 hq
  let N := Fintype.card (BalancedQubitIndex q hq)
  let K := D.logicalQubits
  let dX := lpsDressedZLowerBound q hq
  let dZ := (Fintype.card (EdgeOrbit401 q hq) / 10000000) *
    lpsDressedXLowerBound q
  (D.complex.binaryCSSCode).stabilizer.IsSourceValidBinary ∧
    D.DressedXDistanceAtLeast dX ∧
    D.DressedZDistanceAtLeast dZ ∧
    BoundedLocality
      (QEC.QLDPC.chainComplexBoundaryIncidence D.complex).localityProfile 806 ∧
    N = blocklengthCoefficient * q * (q ^ 2 - 1) ^ 2 ∧
    (q ^ 2 - 1) ^ 2 ≤ K ∧
    K ≤ (201 * (q ^ 2 - 1)) ^ 2 ∧
    N ^ 4 ≤ dimensionLowerCoefficient * K ^ 5 ∧
    K ^ 5 ≤ dimensionUpperCoefficient * N ^ 4 ∧
    N ^ 3 ≤ dressedXPowerCoefficient * dX ^ 5 ∧
    N ^ 3 ≤ dressedZPowerCoefficient * dZ ^ 5
\end{leanlisting}

\subsubsection{Expanding matrices for almost linear lifted product codes}
\label{appendix:almost-input}

Theorem~\ref{thm:almost} takes \lean{PropositionOneExpandingMatrixFamily} as its input.
This is the matrix existence statement corresponding to Proposition 1 of Ref.~\cite{PK20}.
It supplies sparse matrices over a cyclic group algebra, with expansion bounds for both the matrix and its star transpose.
These bounds control the two logical distance sectors.

Let $\ell>1$ be the cyclic lift size and write $G_\ell=\mathbb Z/\ell\mathbb Z$.
There are fixed integers $w\ge4$ and $\gamma>0$ such that, for each $\ell$, one can choose
\begin{equation}
 A_\ell\in\mathbb F_2[G_\ell]^{m_\ell\times ws_\ell},\qquad
 m_\ell=2\lfloor w/4\rfloor s_\ell,\qquad
 \gamma\lfloor\log_2\ell\rfloor\le s_\ell\le\gamma(\lfloor\log_2\ell\rfloor+1).
 \label{eqn:expanding-matrix-size}
\end{equation}
The integer $s_\ell$ controls the dimensions of the matrix over the group algebra.
It is denoted by \lean{n} in Listing~\ref{lst:pk20-premise} and should be distinguished from the physical quantum code length.
Let $T_{A_\ell}$ and $T_{A_\ell^*}$ be the binary linear maps obtained from $A_\ell$ and its star transpose by expanding the group coordinates.
Each map sends every coordinate unit vector to a word of weight at most $w$.

There are also fixed positive rational constants $r_0$ and $\mu$ such that, for either map $T:\mathbb F_2^I\to\mathbb F_2^J$,
\begin{equation}
 \operatorname{wt}(x)\le r_0|I|
 \quad\Longrightarrow\quad
 \operatorname{wt}(Tx)\ge\mu\operatorname{wt}(x)
 \qquad(x\in\mathbb F_2^I).
 \label{eqn:matrix-expansion}
\end{equation}
The sets $I$ and $J$ index the input and output coordinates, respectively.
For $T_{A_\ell}$, we have $|I|=ws_\ell\ell$ and $|J|=m_\ell\ell$; these dimensions are interchanged for $T_{A_\ell^*}$.
In Lean, $r_0$ and $\mu$ are stored as numerator/denominator pairs, and Equation~\eqref{eqn:matrix-expansion} is expressed by cross multiplication.
All four constants $w,\gamma,r_0,\mu$ are chosen before $\ell$ varies.
These hypotheses are the input to the distance argument in Section~\ref{sec:balanced-almost}; the resulting length and distance scales are displayed in Equation~\eqref{eqn:almost-scales}.

\begin{leanlisting}{Expanding matrix assumption used by Theorem~\ref{thm:almost}}{lst:pk20-premise}
-- QEC.Coding.ExpanderCodes.AlmostLinear:41-69
def PropositionOneExpandingMatrixFamily : Prop :=
  ∃ (w gamma radiusNumerator radiusDenominator
      expansionNumerator expansionDenominator : Nat),
    4 ≤ w ∧
    0 < gamma ∧
    0 < radiusNumerator ∧
    0 < radiusDenominator ∧
    0 < expansionNumerator ∧
    0 < expansionDenominator ∧
    ∀ (ell : Nat) [NeZero ell], 1 < ell →
      ∃ (n m : Nat)
        (A : Matrix (Fin m) (Fin (w * n))
          (MonoidAlgebra (ZMod 2) (Cyclic.CyclicGroup ell))),
        m = 2 * (w / 4) * n ∧
        gamma * Nat.log 2 ell ≤ n ∧
        n ≤ gamma * (Nat.log 2 ell + 1) ∧
        (∀ i : Fin (w * n) × Cyclic.CyclicGroup ell,
          HammingWeight (PKDistance.tannerAction A (Pi.single i 1)) ≤ w) ∧
        (∀ i : Fin m × Cyclic.CyclicGroup ell,
          HammingWeight
            (PKDistance.tannerAction (LiftedProduct.starTranspose A)
              (Pi.single i 1)) ≤ w) ∧
        (PKDistance.tannerAction A).SmallSetExpandingAtRelativeRadius
          radiusNumerator radiusDenominator
          expansionNumerator expansionDenominator ∧
        (PKDistance.tannerAction
          (LiftedProduct.starTranspose A)).SmallSetExpandingAtRelativeRadius
            radiusNumerator radiusDenominator
            expansionNumerator expansionDenominator
\end{leanlisting}

\subsection{Resource accounting for constant overhead distillation}
\label{appendix:applications-statements}

Theorem~\ref{thm:constant-overhead} states the output error and overhead guarantees of the distillation protocol.
Listing~\ref{lst:distillation-full} combines that conclusion with the resource conversions and catalyst preparation needed to interpret the resource cost.

Let $r$ index the protocol sequence and let $p$ be the physical local stochastic rate.
For fixed constants $p_0,\Omega>0$, every $p<p_0$ with $p\ge0$ and every target error $\epsilon>0$ admit a starting index $r_{\min}$ such that
\begin{equation}
 r\ge r_{\min}\quad\Longrightarrow\quad
 p_{\rm acc}(r)=1,\qquad \epsilon_{\rm out}(r)\le\epsilon,\qquad
 \frac{N_{\rm in}(r)}{N_{\rm out}(r)}\le\Omega.
 \label{eqn:appendix-distillation-guarantee}
\end{equation}
Here $p_{\rm acc}$ is the acceptance probability.
The input and output counts and the output error are defined in Section~\ref{sec:distillation}.
The guarantee holds for every input preparation with local stochastic rate at most $p$.
In Listing~\ref{lst:distillation-full}, \lean{threshold}, \lean{firstIndex} and \lean{index} correspond to $p_0,r_{\min},r$, respectively.

To express this guarantee in terms of $T$ resources, the protocol uses finite conversions with the resource counts
\begin{equation}
 4b\,T\longrightarrow b\,\mathrm{CCZ},\qquad
 b\,\mathrm{CCZ}+T_{\rm cat}\longrightarrow 2b\,T+T_{\rm cat},
 \label{eqn:resource-conversion}
\end{equation}
where $b\in\mathbb N$ is the batch count, $T$ and $\mathrm{CCZ}$ denote the respective resource states, and $T_{\rm cat}$ is the returned catalyst.
The arrows specify how many resource states each conversion consumes and produces.
The first uses the four-$T$ Toffoli construction~\cite{Jones13}; the second is the catalyzed conversion of a CCZ state into two $T$ states~\cite{GF19}.
The code records the resulting resource counts through \lean{why24RunConversionProgram}.
For $b>0$, combining these conversions with the distillation stage gives
\begin{equation}
 \frac{4bN_{\rm in}(r)}{2bN_{\rm out}(r)}\le3360.
 \label{eqn:converted-overhead}
\end{equation}
The numerator counts the input $T$ resources and the denominator the fresh output $T$ resources; the returned catalyst is accounted for separately.

\begin{samepage}
The catalyst required by this conversion can be prepared with positive success probability and arbitrarily small target error by increasing the distillation depth.
The cost amortization clause uses a specified fixed preparation schedule.
For amortization, take $b\ge1$ batches at protocol index $r=b$.
Writing $c_{\rm seed}>0$ for its certified seed cost and $c_{\rm seed/fresh}(b)$ for the amortized cost function in the theorem, we have
\begin{equation}
 c_{\rm seed/fresh}(b)=\frac{c_{\rm seed}}{2bN_{\rm out}(b)},\qquad
 \lim_{b\to\infty}c_{\rm seed/fresh}(b)=0.
 \label{eqn:catalyst-amortization}
\end{equation}
\end{samepage}
The final \lean{Tendsto} clause of Listing~\ref{lst:distillation-full} expresses this limit.
The preparation clauses use \lean{seedParameter} and allow the schedule to vary with the requested catalyst error.
The displayed limit concerns the fixed seed schedule: its initialization cost per fresh output vanishes as the output count grows.

\begin{leanlisting}{Theorem~\ref{thm:constant-overhead} with resource conversions and catalyst accounting}{lst:distillation-full}
-- QEC.Results.Distillation.ConstantOverhead:372-453
theorem why24_constant_overhead_distillation :
    exists threshold Omega : Real,
      0 < threshold ∧
      0 < Omega ∧
      (forall epsilon : Real, 0 < epsilon ->
        forall physicalRate : Real, 0 <= physicalRate ->
          physicalRate < threshold ->
            exists firstIndex : Nat,
              forall index : Nat, firstIndex <= index ->
                forall preparation : Why24LocalStochasticPreparation
                    (Fin (why24OneRoundInputCount index))
                    (why24IdealCCZRegisterDensity
                      (why24OneRoundInputCount index)),
                  preparation.rate <= physicalRate ->
                    let protocol :=
                      why24OneRoundMagicStateDistillationProtocol index preparation
                    protocol.run = why24OneRoundRun index preparation ∧
                      protocol.inputErrorModel = InputErrorModel.localStochastic
                        (why24LocalStochasticInputErrorModel preparation) ∧
                      protocol.behavior = .activeCorrection (QuantumChannel.id _) ∧
                      protocol.run.resourceCount = 70 * f1024Why24OddLength index ∧
                      why24OneRoundOutputCount index =
                        f1024Why24OddDimension index ∧
                      protocol.run.successProbability = 1 ∧
                      protocol.successfulOutputError <=
                        why24GeometricErrorEnvelope index ∧
                      protocol.successfulOutputError <= epsilon ∧
                      (why24OneRoundInputCount index : Real) /
                        why24OneRoundOutputCount index <= Omega) ∧
      (forall index batch : Nat, 0 < batch ->
        ((4 * batch * why24OneRoundInputCount index : Nat) : Real) /
            (2 * batch * why24OneRoundOutputCount index) <= 3360) ∧
      (forall batch : Nat,
        why24RunConversionProgram (why24FourTToCCZBatch batch)
            { tCount := 4 * batch
              cczCount := 0 } =
            some
              { tCount := 0
                cczCount := batch } ∧
          why24RunConversionProgram (why24CCZCatalyticTBatch batch)
            { tCount := 1
              cczCount := batch } =
            some
              { tCount := 2 * batch + 1
                cczCount := 0 }) ∧
      0 < why24CatalystSeedCost ∧
      (exists seedParameter : ErrorParameter,
        seedParameter.rate = 1 / 200 ∧
        exists hsmall :
            (bh12InputCount 1 : Real) * seedParameter.rate <= 1 / 4,
          exists hhalf : seedParameter.rate <= 1 / 2,
            exists hstrong :
                quadraticCoefficient 1 * seedParameter.rate <= 1 / 2,
              exists schedule : BH12PathScheduleIndex 1,
                why24CatalystSeedCost =
                    bh12PathScheduleTypedSuccessfulOutputCost 1 seedParameter
                      hsmall hhalf hstrong schedule ∧
                  DependentBatch.physicalPolylogarithmicExpectedYieldOverhead
                    (bh12PathScheduleInputSlots 1) (bh12PathScheduleOutputSlots 1)
                    (bh12PathScheduleTargetError 1 seedParameter)
                    (fun schedule =>
                      (bh12PathParameter 1 seedParameter
                        schedule.depth schedule.path).rate)
                    (bh12PathScheduleTypedYieldFactor 1 seedParameter
                      hsmall hhalf hstrong)
                    (bh12BlockExponent 1)
                    (bh12QuadraticEnvelopeCostCoefficient 1 seedParameter) ∧
                  (forall seedEpsilon : Real, 0 < seedEpsilon ->
                    exists depth : Nat, forall path : depthOutputPath 1 depth,
                      0 < rawSuccessMass 1
                          (bh12PathParameter 1 seedParameter depth path) ∧
                        0 < (bh12PathRun 1 seedParameter depth path).successProbability ∧
                          (forall port : OddRow 1,
                            bh12PathSelectedPortTransportedDensity 1 seedParameter
                                depth path port =
                              OneQubit.diagonalAResource
                                (bh12TotalNextParameter 1
                                  (bh12PathParameter 1 seedParameter depth path)
                                  port)) ∧
                            (bh12PathParameter 1 seedParameter depth path).rate <=
                              seedEpsilon)) ∧
      Tendsto why24CatalystSeedCostPerFreshOutput atTop (nhds 0)
\end{leanlisting}
\end{document}